\documentclass[12pt,a4paper]{article}
\usepackage[utf8]{inputenc}
\usepackage{scalerel,stackengine,amsmath}
\usepackage{amsfonts}
\usepackage{amssymb}
\usepackage{hyphenat}
\usepackage{enumitem}
\usepackage{dsfont}
\usepackage[left=2cm,right=2cm,top=2.5cm,bottom=2.5cm]{geometry}
\usepackage{cite}
\usepackage{slashed}
\usepackage{hyperref}
\usepackage{graphicx}
\usepackage{caption}
\usepackage{float}
\usepackage{subcaption}
\usepackage[labelformat=simple]{subcaption}

\usepackage{authblk}
\hypersetup{
	unicode=true,          
	pdftoolbar=true,        
	pdfmenubar=true,        
	pdffitwindow=false,     
	pdfstartview={FitH},    
	pdftitle={QuantumCoherenceUstableSystems},    
	pdfauthor={},     
	pdfsubject={},   
	pdfcreator={},   
	pdfproducer={Producer}, 
    pdfkeywords={Quantum Coherence} {Unstable systems} {non-hermitian Hamiltonian} {density matrix} {Bloch sphere} {meson oscillations},
    pdfnewwindow=true,      
	colorlinks=true,       
	linkcolor=blue,          
	citecolor=red,        
	filecolor=black,      
	urlcolor=blue           
}

\usepackage{ifthen}
\usepackage{tikz}
\usepackage{xspace}
\usepackage{physics,bm}

\newcommand{\bvec}[1]{\vectorbold{#1}\xspace}
\newcommand{\vecE}{\bvec E\xspace}
\newcommand{\vecG}{\bvec \Gamma\xspace}

\newcommand{\vecb}{\bvec b\xspace}
\newcommand{\veca}{\boldsymbol a\xspace}
\newcommand{\absE}{|\vecE|\xspace}

\newcommand{\absb}{|\vecb|\xspace}

\newcommand{\vecPauli}{\pmb \sigma\xspace}
\newcommand{\Heff}{\textrm{H}_{\textrm{eff}}\xspace}
\newcommand{\R}{\mathbb{R}\xspace}
\newcommand{\vecH}{\bvec{H}_{\textrm{eff}}\xspace}
\newcommand{\absH}{|\bvec{H}_{\textrm{eff}}|\xspace}

\newcommand{\ie}{{i.e.}\xspace}

\newcommand{\lrb}[1]{\left( #1 \right)}
\newcommand{\lrsb}[1]{\left[ #1 \right]}

\newcommand{\lrBigcb}[1]{\Big\{ #1 \Big\}}

\renewcommand{\Re}[1]{\operatorname{Re}\lrsb{#1}\xspace}
\renewcommand{\Im}[1]{\operatorname{Im}\lrsb{#1}\xspace}

\renewcommand{\Tr}[1]{\textrm{Tr}\lrsb{#1}\xspace}

\newcounter{NumArgs}

\newcommand{\eqs}[1]{\setcounter{NumArgs}{0}\foreach\i in{#1}{\stepcounter{NumArgs}}%
	\ifthenelse{\equal{\theNumArgs}{1}}{(\ref{#1})}%
	{\ifthenelse{\equal{\theNumArgs}{2}}%
		{\foreach\i[count=\q]in{#1}{\ifthenelse{\equal{\q}{\theNumArgs}}{and (\ref{\i})}{(\ref{\i})~}}}%
		{\foreach\i[count=\q]in{#1}{\ifthenelse{\equal{\q}{\theNumArgs}}{and (\ref{\i})}{(\ref{\i}),~}}}}}

\newcommand{\refs}[1]{\setcounter{NumArgs}{0}\foreach\i in{#1}{\stepcounter{NumArgs}}%
	\ifthenelse{\equal{\theNumArgs}{1}}{(\ref{#1})}%
	{\ifthenelse{\equal{\theNumArgs}{2}}%
		{\foreach\i[count=\q]in{#1}{\ifthenelse{\equal{\q}{\theNumArgs}}{and (\ref{\i})}{(\ref{\i})~}}}%
		{\foreach\i[count=\q]in{#1}{\ifthenelse{\equal{\q}{\theNumArgs}}{and (\ref{\i})}{(\ref{\i}),~}}}}}

\newcommand{\Figs}[1]{\setcounter{NumArgs}{0}\foreach\i in{#1}{\stepcounter{NumArgs}}%
	\ifthenelse{\equal{\theNumArgs}{1}}{Figure~\ref{#1}}%
	{\ifthenelse{\equal{\theNumArgs}{2}}%
		{Figures~\foreach\i[count=\q]in{#1}{\ifthenelse{\equal{\q}{\theNumArgs}}{and \ref{\i}}{\ref{\i}~}}}%
		{Figures~\foreach\i[count=\q]in{#1}{\ifthenelse{\equal{\q}{\theNumArgs}}{and \ref{\i}}{\ref{\i},~}}}}}

 \renewcommand{\theequation}{\arabic{section}.\arabic{equation}}

\usepackage{indentfirst}

\definecolor{ao}{rgb}{0.0, 0.0, 1.0}
\newenvironment{bl}[1]{{\color{ao}  #1}}

\newcommand*{\email}[1]{%
	\footnotesize\href{mailto:#1}{\bl{#1}}
}

\title{\textbf{Quantum Coherence of Critical Unstable Two-Level Systems}}

\author[]{Dimitrios Karamitros\footnote{\email{dimitrios.karamitros@manchester.ac.uk}} }
\author[]{Thomas McKelvey\footnote{\email{thomas.mckelvey@manchester.ac.uk}} }
\author[]{Apostolos Pilaftsis\footnote{\email{apostolos.pilaftsis@manchester.ac.uk}} }

\affil[]{
\normalsize\textit{\hspace{1.4cm}Department
 of Physics and Astronomy, University of Manchester,\newline Manchester, M13 9PL, United Kingdom}
}
 
\date{\empty}

\begin{document}
\setcounter{page}{1}

{\let\newpage\relax\maketitle}

\maketitle

\flushbottom
\vspace{-1cm}
\begin{abstract}
  \noindent 
  We study in detail the dynamics of unstable two-level quantum systems by adopting the Bloch-sphere formalism of qubits. By employing the Bloch-vector representation for such unstable qubit systems, we identify a novel class of critical scenarios in which the so-called energy-level and decay-width vectors, ${\bf E}$ and ${\bf\Gamma}$, are ortho\-gonal to one another, and the parameter $r = |{\bf \Gamma}|/(2|{\bf E}|)$ is less than~1. Most remarkably, we find that critical unstable qubit systems exhibit atypical behaviours like coherence--decoherence oscillations when analysed in an appropriately defined co-decaying frame of the system. In the same frame, a unit Bloch vector ${\bf b}$ describing a pure critical qubit will sweep out {\em unequal} areas during equal intervals of time, while rotating about the vector   ${\bf E}$. These phenomena emerge beyond the usual oscillatory pattern due to the energy-level difference of the two-level quantum system. Interestingly enough, we observe that these new features will persist even for quasi-critical scenarios, in which the vectors ${\bf E}$ and ${\bf\Gamma}$ are not perfectly orthogonal to each other. Applications of our results to quantum information and to unstable meson--antimeson and other systems are discussed. 

\end{abstract}

{\small {\sc Keywords:} Quantum Coherence; Unstable Qubits; Density Matrix; Effective Hamiltonian.}
\newpage
\tableofcontents
\newpage

\section{Introduction}\label{sec:Intro}
\setcounter{equation}{0}

Quantum entanglement and quantum coherence are two topics in Quantum Mechanics~(QM) that received much attention in several modern physics explorations. They have extensively been tested in terrestrial experiments, and suggestions have been made recently to probe them in satellite missions in space~\cite{Alonso:2022oot}. These two concepts of~QM play also a central r\^{o}le in our understanding of quantum systems, including elementary systems such as two-level quantum systems, also known as {\it qubits}~\cite{PhysRevA.51.2738}.

Historically, two-level quantum systems were introduced in 1937 by Rabi~\cite{Rabi:1937dgo} to  describe his nominal oscillatory phenomena such as spin precession of fermions in the presence of magnetic fields, with  notable applications to nuclear magnetic resonance and medicine. In the context of particle physics, these ideas have been reinvented by Lee, Oehme and Yang (LOY)~\cite{Lee:1957qq} to address possible phenomena of charge (C) and charge-parity (CP) violation in the kaon system through $K^0\bar{K}^0$ oscillations.  Subsequently, Pontecorvo~\cite{Pontecorvo:1957cp,Pontecorvo:1967fh}, and Maki, Nakagawa and Sakata~\cite{Maki:1962mu} suggested that similar phenomena of oscillations of flavour and charge may take place with neutrinos, if the latter happen to have a non-zero mass as confirmed later by experiment. 

If the two-level quantum system turns out to be unstable like the kaon system mentioned above,  one has then to consider decay-width effects in addition to oscillations. These effects have been taken   into account by LOY~\cite{Lee:1957qq} who made use of the famous
effective approximation developed in the 1930s by Weisskopf and Wigner~(WW)~\cite{Weisskopf1930}. Although the work by WW  was originally aimed at approximating the behaviour of decaying spectral lines in atoms, it has eventually found a wide range of applicability in the realm of quantum theory. In the WW approximation, the dynamics of an unstable quantum system, such as an unstable qubit, may be conveniently described by making use of an effective non-Hermitian Hamiltonian, $\textrm{H}_\textrm{eff}$.
The time evolution of an unstable quantum state, $\ket{\Psi}$, is then governed by an effective Schr\"{o}dinger equation,
\begin{equation}
    i\partial_t \ket{\Psi} = \textrm{H}_\textrm{eff} \ket{\Psi},
    \label{eq:Schrodinger_def}
\end{equation}
which can be used to compute observable quantities related to unstable qubits, like kaon, $B$- and $D$-meson systems~\cite{ParticleDataGroup:2022pth}.

At sufficiently large times, an initially mixed unstable qubit is expected to approach a pure QM state aligned to its long-lived eigenstate of ${\rm H}_{\rm eff}$, since its complementary short-lived eigenstate must have already decayed away. However, if the two eigenstates of ${\rm H}_{\rm eff}$ happen to have equal lifetimes, the evolution of quantum coherence for the unstable qubit will become rather complex. In order to accurately capture all the features emanating from such dynamics, we will adopt in our study the Bloch-sphere formalism of qubits~\cite{PhysRev.70.460}. In particular, we will employ the Bloch vector representation to decompose the non-Hermitian Hamiltonian ${\rm H}_{\rm eff}$ in terms of the energy-level and decay-width four-vectors, $E^\mu = (E^0, {\bf E})$ and $\Gamma^\mu = (\Gamma^0,{\bf\Gamma})$, respectively. In this Bloch-sphere formulation, the degree of quantum coherence is determined by an analogous decomposition of the density matrix $\rho$ describing the qubit, in terms of the so-called Bloch four-vector: $a^\mu = (a^0, {\bf a})$. For an unstable qubit, it proves more convenient to define the {\em normalised} density matrix $\hat{\rho} \equiv\rho / \text{Tr}\,\rho$, such that $\text{Tr}[\hat{\rho} (t)] = 1$ at all times, $t$. This normalisation defines some sort of a {\em co-decaying} frame, in which the {\em normalised} density matrix $\hat{\rho}$ can be equivalently expressed by the normalised Bloch four-vector: $b^\mu = (1\,, {\bf b})$.
A pure unstable qubit will be characterised by having a Bloch three-vector~${\bf b}$ of unit norm, i.e.~$|{\bf b}| =1$. Instead, a fully mixed unstable qubit possesses a vanishing Bloch three-vector, ${\bf b} = {\bf 0}$.

In this paper we identify a new class of {\em critical} unstable two-level  systems, in which the energy-level and decay-width vectors, ${\bf E}$ and ${\bf\Gamma}$, are orthogonal to one another, and their norm ratio, $r = |{\bf \Gamma}|/(2|{\bf E}|)$, turns out to be less than~1\footnote{The limiting scenario $r = 1$ was discussed before in~\cite{Pilaftsis:1997dr}, and corresponds to a non-diagonalisable $\textrm{H}_{\rm eff}$.}. Specifically, we find that these critical unstable qubits display a number of unusual quantum properties. Most strikingly, if the critical unstable\- qubit is prepared to be initially at a fully mixed state (with ${\bf b} = {\bf 0}$), they will exhibit coherence--decoherence oscillations when viewed from the aforementioned co-decaying frame, which is described by the normalised density matrix $\hat{\rho}$ of the system. 
Moreover, if the critical unstable qubit is initially in a 
pure QM state given by a normalised unit Bloch vector, with   
$|{\bf b}| = 1$ and ${\bf b} \nparallel {\bf E}$, then the
unit vector ${\bf b}(t)$ will rotate about the direction defined by the energy-level vector~${\bf E}$, and it will sweep out {\em unequal} areas in equal amounts of time. We note here that these phenomena differ from the usual oscillatory pattern that would result from the energy-level difference of the two-level quantum system. In fact, our analysis shows that these new features will persist even for quasi-critical scenarios, in which ${\bf E}$ is not exactly perpendicular to ${\bf\Gamma}$. In~this context, we will show how our results can be applied to unstable  $K^0\bar{K}^0$, $B^0\bar{B}^0$ and $D^0\bar{D}^0$ meson--antimeson systems. 

In this article, we also study the impact of decoherence phenomena on the time evolution of the Bloch vector ${\bf b}$ in both critical and quasi-critical scenarios. In this case, the Bloch vector~${\bf b}$ will freeze in a direction that crucially depends on the dimensionless decoherence parameter, $\zeta = 4|{\bf D}|^2/|{\bf \Gamma}|$, where ${\bf D}$ is the decoherence vector resulting from a Bloch-sphere decomposition.  Thus, we find that critical unstable qubits exhibit unusual quantum behaviours that have no analogue in classical systems. Their simulation would require the use of quantum computers to probe the validity of their many unexpected predictions that will be presented in this work.

The paper is laid out as follows. After this introductory section,  Section~\ref{sec:density_matrix} outlines the density matrix formalism and the WW approximation, thereby laying the groundwork for the upcoming analysis. In Section~\ref{sec:2Level}, we apply the Bloch-sphere formalism to unstable two-level quantum systems, which we call them interchangeably unstable qubits. In the same section, we analyse the asymptotic behaviour of the unstable two-level systems, and deduce the experimental values for  key kinematic parameters that describe the $K$-, $B$- and $D$-meson systems within the context of the Bloch-sphere formalism. Furthermore, we present a novel class of unstable qubits which exhibit atypical asymptotic behaviours, including coherence--decoherence oscillations as observed by considerations of the entanglement entropy. We have called such quantum systems {\em critical} unstable qubits. In Section~\ref{sec:CritScen}, we investigate in depth the dynamics of such critical unstable qubits, and highlight the main features that are unique to this particular class of quantum systems. We also study how these features get affected by the presence of decoherence effects. Section~\ref{sec:QuasiCrit} discusses the dynamics of quasi-critical qubits where the energy-level vector ${\bf E}$ is not exactly perpendicular to the decay-width vector ${\bf \Gamma}$.   Finally, Section~\ref{sec:Concl} summarises the main results of our study and
discusses further research directions. Several technical details are relegated to Appendices~\ref{app:Analytic} and~\ref{app:Nonlinear}.

\section{The Density Matrix}\label{sec:density_matrix}
\setcounter{equation}{0}

The density matrix formalism \cite{Shankar:102017,Sakurai:1341875} is a well-known robust approach to describing the coherent properties of quantum systems. The central object is the 
density-matrix (operator)~$\rho$, which is defined as
\begin{equation}\label{eq:DenMatDef}
    \rho = \sum_{k} w_k \ket{\Psi^k}\bra{\Psi^k} \;,
\end{equation}
where $\ket{\Psi^k}$ are taken here to be normalised states, and $w_k$ is the probability of the system to be in state~$\ket{\Psi^k}$.

The density matrix allows us to directly distinguish between pure and mixed states. A pure state is defined as
\begin{equation}\label{eq:DenMatPureDef}
    \rho = \ket{\Psi}\bra{\Psi} \;,
\end{equation}
which obeys
\begin{equation}
   \label{eq:rho2}
    \rho^2 = \rho\; .
\end{equation}
Instead, mixed states are characterised by the fact that they do not satisfy condition~\eqref{eq:rho2}. Hence, we regard pure states to be fully coherent with vanishing entropy, while mixed states are incoherent, for which the entropy takes non-zero values (see our discussion in Section~\ref{sec:Entropy}). In fact, as we will see in the next section, mixed states may possess some degree of coherence that can be quantified by the magnitude of the Bloch vector ${\bf a}$. 

The density-matrix formalism has the advantage of being able to simultaneously track both the probabilities associated with the states of the system and the correlations between them. Thus, the sum over the diagonal elements should be unitary
\begin{equation}
    \textrm{Tr}\,\rho\, =\, \sum_k \mathbb{P}(k)\, =\,  1 \;,
\end{equation}
where $\mathbb{P}(k)$ is the probability associated with finding the system in the $k$-th state. 

\subsection{Time Evolution}
The density-matrix formalism provides meaningful and consistent results when applied to a simple unitary model, as described by a Hermitian Hamiltonian. However, for an unstable quantum system, an effective description by means of a non-unitary time evolution operator~\cite{Kabir_1996} may obscure the statistical interpretation of the density matrix.

As was discussed in the introduction, the WW approximation may be utilised in the context of an effective Hamiltonian \cite{Lee:1957qq}, under which the time evolution of states follow the Schr\"{o}dinger equation \eqs{eq:Schrodinger_def}, with
\begin{equation}
    \Heff\: =\: \textrm{E}\, -\, \frac{i}{2}\Gamma\;.
    \label{eq:Heff_def}
\end{equation}
In the above expression, $\textrm{E}$ and $\Gamma$ are both Hermitian matrices. For example, $\textrm{E}$ may correspond to the radiatively corrected mass matrix, and $\Gamma$ to the absorptive part of the self-energy of a particle system~\cite{Pilaftsis:1997dr}. One may then show that the density matrix as defined in (\ref{eq:DenMatDef}) follows an equation of motion involving commutator, $[\cdot,\cdot]$, and anti-commutator, $\{\cdot , \cdot \}$, terms, viz.
\begin{equation}
    \frac{d\rho}{dt}\, =\, -i \left(\textrm{H}_{\textrm{eff}} \,\rho - \rho \, \textrm{H}_{\textrm{eff}}^\dagger\right)\, =\, -i \left[\textrm{E} , \rho \right] - \frac{1}{2}\left\{\Gamma, \rho \right\}.
    \label{eq:rho_evolution}
\end{equation}
An exact solution to the linear differential equation~\eqref{eq:rho_evolution} can be given with the help of the non-unitary time-evolution operator, $U(t) = e^{-i\textrm{H}_\textrm{eff} t}$, as follows:
\begin{equation}
    \rho(t)\: =\: e^{-i\textrm{H}_\textrm{eff} t}\, \rho(0)\:  e^{i\textrm{H}_\textrm{eff}^\dagger t}\;.
    \label{eq:rho_evolution_operator}
\end{equation}
By taking the trace, $\Tr{\rho(t)}$, in~\eqs{eq:rho_evolution_operator}, we see that the total probability is not conserved, i.e.
\begin{equation}
    \frac{d}{dt}\textrm{Tr}\,\rho\, =\, \frac{d}{dt} \sum_k \mathbb{P}(k)\, =\, - \textrm{Tr}\big(\Gamma\rho\big)\ \neq\ 0\;.
    \label{eq:Tr_rho_evolution}
\end{equation}
This result is a consequence of the fact that for unstable quantum systems~\cite{Kabir_1996}, only part of the full Hilbert space is 
considered. 

Let us define a sub-space, $\mathcal{H}_1$, of the full Hilbert space, $\mathcal{H}$, which contains all the states at time $t=0$. When these states decay, they decay to another state in the wider Hilbert space not contained within $\mathcal{H}_1$. We may then study the states which remain within $\mathcal{H}_1$ by ensuring that the conditional probabilities we calculate accurately describe the remaining states.  In this way, we find
\begin{equation}
    \mathbb{P}(k | \mathcal{H}_1)\, =\, \frac{\mathbb{P}(k)}{\mathbb{P}(\mathcal{H}_1)}\, =\, \frac{\mathbb{P}(k)}{\sum_{q \in \mathcal{H}_1} \mathbb{P}(q)}\, =\, \frac{\mathbb{P}(k)}{\textrm{Tr}\,\rho}\;,
    \label{eq:rho_norm_def}
\end{equation}
where $\mathbb{P}(k|\mathcal{H}_1)$ is the conditional probability of observing the state $k$ given this state must have come from $\mathcal{H}_1$. This finding allows us to appropriately define the normalised density matrix by dividing the density matrix by its trace,
\begin{equation}\label{eq:NormDenMat}
    \hat{\rho}\, \equiv\, \frac{\rho}{\textrm{Tr}\,\rho}\; ,
\end{equation}
which contains all the conditional probabilities and correlations between states of $\mathcal{H}_1$. We refer to the basis normalised to $\mathcal{H}_1$ as the \textit{co-decaying basis},
or the {\it co-decaying frame}. We may then derive an equation of motion for the normalised density matrix
\begin{equation}
   \label{eq:drhohatdt}
    \frac{d\hat{\rho}}{dt}\ =\ \frac{1}{\textrm{Tr}\,\rho}\,\frac{d\rho}{dt}\: -\:\frac{\rho}{\textrm{Tr}^2\rho}\, \frac{d\,\textrm{Tr}\,\rho}{dt}\ =\ -i \left[\textrm{E} , \hat{\rho} \right] - \frac{1}{2}\left\{\Gamma, \hat{\rho} \right\} +\textrm{Tr}\big(\Gamma\hat{\rho}\big) \, \hat{\rho}\; .
\end{equation}
This adjusted equation of motion may be verified to preserve the condition $\textrm{Tr}\,\hat{\rho} = 1$, implying conserved conditional probability. Obviously, this comes at the cost of introducing a non-linear term into the differential equation~\eqref{eq:drhohatdt}. But as we will see in the next sections, several unexpected and interesting features of quantum coherence become manifest for unstable qubits in the  co-decaying frame as defined above.

\section{Two-Level Quantum Systems}\label{sec:2Level}
\setcounter{equation}{0}

It is known that the density matrix of an $N$-level quantum system possesses a rich geometric structure. It forms  a manifold generated by the coset space: $\textrm{U}(N)/\textrm{U}(N-1) \times \textrm{U}(1)$ \cite{Goyal_2016}. In~the case of the two-level system ($N=2$), we have the homeomorphism: $\text{SU}(2)/\text{U}(1)\sim S^2$, and as such, the generated manifold may be represented by the two-sphere $S^2$, which is now better known as the Bloch sphere. This representation of the density matrix was initially used by Poincar\'{e} in his study of polarisation \cite{PoincareH1889,stokes_2009}, and then adopted by Bloch to study magnetic moments within nuclei \cite{PhysRev.70.460}. It was shown that for the two-level system, the density matrix could be neatly represented through a vector $\veca$, the so-called Bloch vector, when $\rho$ is written in the Pauli basis $\boldsymbol{\sigma} = (\sigma_1\,, \sigma_2\,, \sigma_3)$,
\begin{equation}
    \rho\, =\, \frac{1}{2}\, \Big(\mathds{1} + \veca\cdot\boldsymbol{\sigma} \Big)\,.
\end{equation}
This expression satisfies the core properties of the density matrix, including $\textrm{Tr}\,\rho = 1$, as well as $\rho^2 = \rho$ for pure states  when $|\veca| = 1$. 
For non-unitary (decaying) systems, the expression for $\rho$ must be extended by including a zeroth component, $a^0$, which acts as a normalisation for the density matrix, i.e.
\begin{equation}
    \rho\, =\, \frac{1}{2}\, \Big( a^0 \mathds{1} + \veca\cdot\boldsymbol{\sigma} \Big)\,.
\end{equation}
This last form of the density matrix $\rho$ will help us to systematically investigate the dynamics of the unstable two-level system, through the evolution of the Bloch four-vector: $a^\mu = (a^0 , \veca)$.

\subsection{General Cases}
Let us first consider a general qubit system, before we look closer at specific cases of interest. We may write the density matrix, $\rho$, and the effective Hamiltonian, ${\rm H}_{\rm eff}$, 
defined in~\eqref{eq:Heff_def} as expansions in the Pauli basis as follows:
\begin{equation}
    \rho = \frac{1}{2} a_\mu \overline{\sigma}^\mu, \qquad E = E_\mu \sigma^\mu, \qquad \Gamma = \Gamma_\mu \sigma^\mu,
    \label{eq:Bloch_dec}
\end{equation}
where  $a^\mu = (a^0 , \veca)$, $\sigma^\mu = (\mathds{1}, \boldsymbol{\sigma})$, $\overline{\sigma}^\mu = (\mathds{1},- \boldsymbol{\sigma})$, $E^\mu = (E^0 , \mathbf{E})$ and $\Gamma^\mu = (\Gamma^0, \boldsymbol{\Gamma})$. Substituting  these expressions into the time-evolution equation~\eqref{eq:rho_evolution} for $\rho(t)$ produces a set of coupled differential equations for $a^\mu$,
\begin{subequations}
\begin{equation}
    \frac{da^0}{dt} = - \Gamma_\mu a^\mu,
    \label{eq:a0_eom}
\end{equation}
\begin{equation}
    \frac{d\veca}{dt} = - 2\,\mathbf{E}\times \veca + \boldsymbol{\Gamma}a^0 - \Gamma^0 \veca.
    \label{eq:ai_eom}
\end{equation}
\end{subequations}
Notice that $E^0$ does not contribute since the equation of motion \eqs{eq:rho_evolution} is invariant under the shift: $E \to E + \kappa \, \sigma^0$, where $\kappa$ is an arbitrary constant. The solutions to \eqs{eq:a0_eom,eq:ai_eom}, obtained by evolving $\rho(0)$ according to~\eqref{eq:rho_evolution_operator},  are given in \eqs{eq:a_solution}. 

 The density matrix in the co-decaying frame, $\hat{\rho}$, can be decomposed as
\begin{equation}
    \hat\rho\, =\, \dfrac{1}{2}\, \Big( \mathds{1} + \vecb \cdot \vecPauli \Big)\;,
    \label{eq:normalised_rho_Bloch}
\end{equation}
where the so-called {\it co-decaying} Bloch vector $\mathbf{b} = \veca/a^0$ satisfies  the {\em non-linear} equation of motion,
\begin{equation}
    \frac{d\mathbf{b}}{dt}\, =\, -2\,\mathbf{E} \times\mathbf{b} + \boldsymbol{\Gamma} - (\boldsymbol{\Gamma}\cdot\mathbf{b})\,\mathbf{b}.
\end{equation}
The equation of motion for the co-decaying Bloch vector~${\bf b}$ can be solved using \eqs{eq:a_solution}, from which the complete time evolution of the unstable qubit system can be determined. However, such an expression proves difficult to work with and extract information from, as far as the coherence properties of the qubit are concerned. Nevertheless, in the asymptotic time limit $t \to \infty$, this solution often freezes into a particular\- direction that depends on the input parameters of the model and usually corresponds to the longest lived eigenstate of the effective Hamiltonian~${\rm H}_{\rm eff}$. Under such circumstances, it is then possible to extract analytic expressions for the asymptotic limit of the co-decaying Bloch vector~${\bf b}$, by identifying such points as stable fixed points of the dynamical system. 

To simplify the analysis of two-level systems, we introduce the two dimensionless parameters
\begin{equation}\label{eq:Params}
     \tau\, =\, |\boldsymbol{\Gamma}|\, t, \qquad r\, =\, \frac{|\boldsymbol{\Gamma}|}{2\,|\mathbf{E}|} \; ,
\end{equation}
as well as the unit vectors $\mathbf{e} = \mathbf{E}/|\mathbf{E}|$ and $\boldsymbol{\gamma} = \boldsymbol{\Gamma}/|\boldsymbol{\Gamma}|$. Under this new parameterisation, the time-evolution equation for the co-decaying Bloch vector is given by
\begin{equation}
   \label{eq:dbdt}
    \frac{d\mathbf{b}}{d\tau}\: =\: -\frac{1}{r} (\mathbf{e} \times\mathbf{b})\, +\, \boldsymbol{\gamma}\, -\, (\boldsymbol{\gamma}\cdot\mathbf{b})\,\mathbf{b}\,.
\end{equation}
Taking the inner product on both sides of~\eqref{eq:dbdt} with ${\bf b}$, it is not difficult to derive that the magnitude of the co-decaying Bloch vector will evolve as
\begin{equation}
    \frac{d|\mathbf{b}|^2}{d\tau}\: =\: 2(\boldsymbol{\gamma}\cdot\mathbf{b})\,\big(1 - |\mathbf{b}|^2 \big)\,.
\end{equation}
This last differential equation suggests that the unstable qubit system will tend towards a pure state $|\mathbf{b}| = 1$, provided asymptotically stable solutions exist for the co-decaying Bloch vector, even if the system starts initially as an entirely mixed state, with ${\bf b}(0) = {\bf 0}$. As a result, it is necessary to identify scenarios under which these stable solutions manifest, or more interestingly, when these solutions do not exist.

\begin{figure}[t!]
    \centering
    \includegraphics[width=0.8\linewidth]{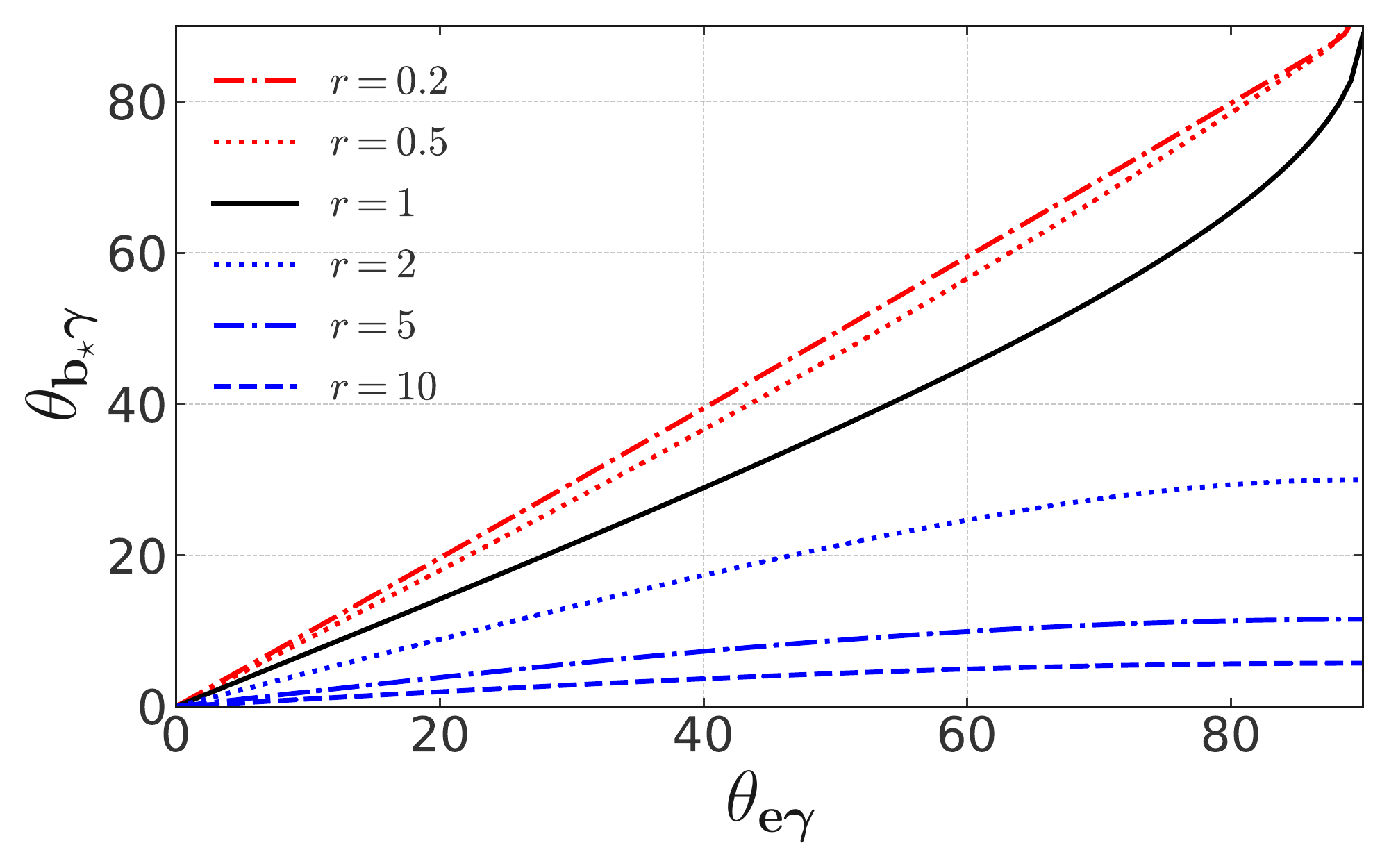}
    \caption{The angle, in degrees, between $\mathbf{b}_\star$ and $\boldsymbol{\gamma}$ as a function of the angle between $\mathbf{e}$ and $\boldsymbol{\gamma}$, for varying values of $r$ shown with different colours.}
    \label{fig:FPAngles}
\end{figure}

As mentioned above, this asymptotic solution usually lies along the longest lived state contained in the density matrix. We may extract this state by constructing a basis from the two unit vectors $\mathbf{e}$ and $\boldsymbol{\gamma}$, upon which we represent the asymptotic solution for the co-decaying Bloch vector
\begin{subequations}\label{eq:FPN}
\begin{equation}
    \mathbf{b}_\star\, \equiv\, \mathbf{b}(\tau \to \infty)\, =\, \alpha \mathbf{e} - \frac{1}{s_\gamma^2 r}(1-\alpha^2) \, \mathbf{e}\times\boldsymbol{\gamma}  -  \frac{c_\gamma}{s_\gamma^2 \alpha}(1-\alpha^2) \, \mathbf{e} \times (\mathbf{e} \times\boldsymbol{\gamma})\,,
\end{equation}
with
\begin{equation}
    \alpha = \left\{\begin{array}{lr}
        +\,\frac{\displaystyle 1}{\displaystyle \sqrt{2}}\,\sqrt{\left(1-r^2\right) + \sqrt{(1-r^2)^2 + 4r^2 c_\gamma^2}}\;, & \text{for } c_\gamma > 0\,,\\[5mm]
        -\,\frac{\displaystyle 1}{\displaystyle \sqrt{2}}\,\sqrt{\left(1-r^2\right) + \sqrt{(1-r^2)^2 + 4r^2 c_\gamma^2}}\;, & \text{for } c_\gamma < 0\,.
        \end{array}\right.
\end{equation}
\end{subequations}
In the above, $s_\gamma = \sin \theta_{\mathbf{e}\boldsymbol{\gamma}}$ and $c_\gamma = \cos \theta_{\mathbf{e}\boldsymbol{\gamma}}$ are the sines and cosines of $\theta_{\mathbf{e}\boldsymbol{\gamma}}$, respectively, where $\theta_{\mathbf{e}\boldsymbol{\gamma}} = 
\angle(\mathbf{e},\boldsymbol{\gamma})$ is the angle between $\mathbf{e}$ and $\boldsymbol{\gamma}$, taking values from the domain $[0^\circ, 360^\circ)$.
A~detailed analysis regarding the derivation of the stable solution is presented in Appendix~\ref{app:Nonlinear}.

Figure~\ref{fig:FPAngles} shows the set of asymptotic solutions by taking the angle between $\mathbf{b}_\star$ and $\boldsymbol{\gamma}$ for all input angles $\theta_{\mathbf{e}\boldsymbol{\gamma}}$. This figure was generated by first solving numerically the equation of motion for the co-decaying Bloch vector ${\bf b}$ in~\eqref{eq:dbdt} for a range of values for $r$ and $\theta_{\mathbf{e}\boldsymbol{\gamma}}$, and then extracting the relevant angle~$\theta_{\mathbf{b}_\star\boldsymbol{\gamma}}$. These results may be verified by calculating the expected angle, $\theta_{\mathbf{b}_\star \boldsymbol{\gamma}}$, through
\begin{equation}
    \cos\theta_{\mathbf{b}_\star\boldsymbol{\gamma}}\, =\, \boldsymbol{\gamma} \cdot \mathbf{b}_\star\, =\, \frac{c_\gamma}{\alpha}\,.
\end{equation}
Then, after carefully calculating relevant limits of $r$, we obtain
\begin{equation}
    \lim_{r \to 0} \theta_{\mathbf{b}_\star\boldsymbol{\gamma}} = \theta_{\mathbf{e}\boldsymbol{\gamma}}, \qquad \lim_{r \to \infty} \theta_{\mathbf{b}_\star\boldsymbol{\gamma}} = 0\,.
\end{equation}
Here, we must caution the reader that the exact value at $r=0$ is undefined, since then we would return to the well-known Rabi rotations of the Bloch vector ${\bf b}(t)$ about the three-vector~${\bf E}$.

It may be verified that the fixed point solution given in (\ref{eq:FPN}) holds for almost any set of input parameters one may choose. However, as may be seen in Appendix~\ref{app:Nonlinear}, when we take $c_\gamma = 0$, the fixed point solution is not guaranteed to be stable, since the two key mathematical quantities evaluated in (\ref{eq:FPCond}) vanish. This specific circumstance corresponds to systems where the angle between $\mathbf{e}$ and $\boldsymbol{\gamma}$ is $\pm\pi/2$, \textit{i.e.} the two vectors are perpendicular. If we repeat this analysis again for the perpendicular system, we see that the Bloch space may be spanned by the basis $\left\{\mathbf{e}, \boldsymbol{\gamma}, \mathbf{e} \times\boldsymbol{\gamma} \right\}$, and the fixed points are given by
\begin{equation}\label{eq:CSFP}
    \mathbf{b}_\star\: =\: \frac{\sqrt{r^2-1}}{r} \boldsymbol{\gamma}\, -\, \frac{1}{r} \mathbf{e} \times \boldsymbol{\gamma}.
\end{equation}
The dynamical system does not have stable fixed points when $r<1$, and $\mathbf{e}$ is perpendicular to $\boldsymbol{\gamma}$, as this would result in $\vecb_*$ being complex. Consequently, we observe a Hopf bifurcation of the perpendicular system at $r=1$. For an effective Hamiltonian ${\rm H}_{\rm eff}$ that respects CPT invariance, the location of this bifurcation would correspond to a two-level system having a ${\rm H}_{\rm eff}$ in Jordan form. Such effective Hamiltonians are completely degenerate, and are non-diagonalisable~\cite{Pilaftsis:1997dr}. 

In view of the findings from the above analysis, we may distinguish the special class of systems, in which $\mathbf{e}\perp \boldsymbol{\gamma}$
and $r<1$, from ordinary unstable qubits, and characterise them as \textit{critical} unstable qubit systems. These critical unstable two-level systems will be studied in more detail in Section~\ref{sec:CritScen}.

\subsection{Application to CP Violating Systems}\label{sec:Meson}

The methodology that we have been developing can now find applications to CP-violating decaying systems described by CPT-invariant effective Hamiltonians ${\rm H}_{\rm eff}$ similar to \eqs{eq:Heff_def}, with $E_{21}=E_{12}^{*}$, $E_{11}=E_{22} \in \mathbb{R}$, $\Gamma_{11}=\Gamma_{22} \in \mathbb{R}$, and $\Gamma_{21}=\Gamma_{12}^{*}\in \mathbb{C}$. The elements of the Hamiltonian can be expressed in terms the Bloch-decomposed parameters as
\begin{align}
    &\vecE\, =\, \lrb{ -\Re{E_{12}}, \, \Im{E_{12}},0}\,, \nonumber\\ 
    &\vecG\, =\, \lrb{ -\Re{\Gamma_{12}}, \, \Im{\Gamma_{12}},0}\,, \label{eq:Bloch_EG_meson}\\
    &E^0\, =\, E_{11}\, =\, E_{22}\,,\nonumber\\ 
    &\Gamma^0\, =\, \Gamma_{11}\, =\, \Gamma_{22}\;. \nonumber
\end{align}
%

The effective Hamiltonian of this system, ${\rm H}_{\rm eff}$, can be diagonalised by means of a similarity trans\-formation, $X^{-1} \, {\rm H}_{\rm eff} \, X$, which rotates the system from the flavour basis, $\ket{0}$ and $\ket{\bar 0}$, to the mass eigenstate basis, $\ket{\pm}$~\footnote{A brief review of the formalism presented here may be found in~\cite{Kabir:1989dd, ParticleDataGroup:2022pth}.}. Defining the transformation matrix, $X$, as
\begin{equation}
    X\, =\, \lrb{\begin{array}{cc}
         p&p  \\
         -q&q 
    \end{array}} \;,
    \label{eq:pq_matrix}
\end{equation}
with $|p|^2 + |q|^2 = 1$, we find the useful relation,
\begin{equation}
\left|\dfrac{q}{p}\right|^2 =\ \left|\frac{E_{12}^* - \frac{i}{2} \Gamma_{12}^*}{E_{12} - \frac{i}{2} \Gamma_{12}}\right|\:
    =\, \,\dfrac{1+r^2 - 2r\sin \theta_{\mathbf{e}\boldsymbol{\gamma}}}{\sqrt{1+r^4 + 2r^2 \cos 2\theta_{\mathbf{e}\boldsymbol{\gamma}}}} \; .
    \label{eq:pq_ratio}
\end{equation}


\begin{table}[t!]
    \centering
    \begin{tabular}{|c||c|c|c|}
        \hline
        Meson & $r$ &  $\theta_{\mathbf{e}\boldsymbol{\gamma}} [^\circ]$  & $|\mathbf{E}| [\textrm{ps}^{-1}]$\\[0.1cm]
        \hline\hline
        $K^0$  & $1.05315 \pm 4.5 \times 10^{-4}$ &  $179.6322 \pm   6 \times 10^{-4}$ & $5.293 \times 10^{-3} \pm 2.021\times10^{-10}$\\[0.1cm]
        \hline
        $D^0$&  $1.511 \pm 0.147$ & $179.3 \pm  2.0$ & $10^{-2} \pm 4\times 10^{-6}$\\[0.1cm]
        \hline
        $B_{\rm d}^0$&  $(8.3\pm 8.1) \times 10^{-3}$ & $270 \pm 90$ & $0.5056 \pm 4 \times 10^{-7} $\\
        \hline
        $B_{\rm s}^0$ &  $(2.62 \pm 0.38) \times 10^{-3}$ & $185.77 \pm 31.53$ & $17.77 \pm 10^{-5}$\\[0.1cm]
        \hline
    \end{tabular}
    \caption{Meson Oscillation data represented as $r$, $\theta_{\mathbf{e}\boldsymbol{\gamma}}$, and $|\mathbf{E}|$ which reproduce meson oscillation experimental data.
    In order to obtain these results, we consider the parameters $|q/p|$ and $\Delta \Gamma$ within $1\sigma$ error with $\Delta m$ taken to its central value.
    The experimental values for the various CP-violating parameters for $K^0$ and $D^0$ are found in~\cite{ParticleDataGroup:2022pth}, while for $B_{\rm d,s}$ in~\cite{HFLAV:2022pwe}. 
    }
    \label{tab:MesonData}
\end{table}

Upon diagonalization of the effective Hamiltonian~${\rm H}_{\rm eff}$, one usually obtains the two eigenvalues
\begin{equation}
    \lambda_{\pm} =  E^0 - \dfrac{i}{2} \, \Gamma^0 \ \pm \absH \;.
    \label{eq:CP_eigenvalues}
\end{equation}
In the above, we have defined the effective Hamiltonian three-vector: $\vecH = \vecE - \dfrac{i}{2} \vecG$, whose  magnitude should be understood as $\absH = \sqrt{\vecH \cdot \vecH}\, \in\,\mathbb{C}$. The real and imaginary parts of the eigenvalues correspond to the masses and decay widths of the CP-violating system, respectively, which are
\begin{align}
    m_{\pm}\, =\, E^0 \pm \Re{ \absH }\,,\qquad 
    \Gamma_{\pm}\, =\, \Gamma^0 \mp 2 \Im{ \absH} \;.
    \label{eq:masses_widths}
\end{align}
We will call the eigenstate with mass $m_-$ ($m_+$)  {\em short-lived} ({\em long-lived}), with mass and width differences given by
\begin{eqnarray}
    \Delta m\, \equiv\, m_+ - m_-\, =\, 2\,\Re{ \absH }\,,\qquad 
    \Delta \Gamma\, \equiv\, \Gamma_- - \Gamma_+\, =\, 4\,\Im{ \absH } \;.
    \label{eq:mass_width_diffs}
\end{eqnarray}
%

The so-called right mass eigenstates of ${\rm H}_{\rm eff}$, denoted below as $\ket{\pm}$,  do not form an orthogonal basis in Hilbert space. Therefore, it is convenient to express these in the original orthonormal flavour basis, $\ket{0}$ and $\ket{\bar 0}$, using the transformation matrix~\eqs{eq:pq_matrix}, 
\begin{equation}
    \ket{\pm}\: =\: p\, \ket{0}\, \pm\, q \, \ket{\bar 0}   \;.
    \label{eq:m_states_from_flavour}
\end{equation}
In this basis, the density matrix $\rho_0 = \rho (0)$ at $t=0$ reads
\begin{equation}
    \rho_{0} = w_{0} \ket{0}\bra{0} + w_{\bar 0} \ket{\bar 0}\bra{\bar 0} + w_{\rm I}\ket{0}\bra{\bar 0}+ w_{\rm I}^*\ket{\bar 0}\bra{0} \;,
    \label{eq:rho0_mass_basis}
\end{equation}
with $w_{0}, \, w_{\bar 0} \in \R$, and $w_{\rm I} \in \mathbb{C}$. The trace of an operator $\widehat O$ over this basis is defined as
\begin{equation}
    \text{Tr}\,\widehat O \, =\, \bra{0} \widehat O \ket{0} + \bra{\bar 0} \widehat O \ket{\bar 0}\;.
    \label{eq:CP_trace_def}
\end{equation}
This allows us to normalise the initial state as $\text{Tr}\,\rho_0 = 1$, \ie $w_{\bar 0} = 1- w_0$. Notice that the initial state is exactly pure when one has $w_0 = \big(\,1 \pm \sqrt{1-4|w_I|^2}\,\big)/2$.

The time evolution of the density matrix is dictated by \eqs{eq:rho_evolution_operator}. In the mass basis, $\rho(t)$ becomes
\begin{align}
    \rho(t) \: =\: 
    &\ w_+ \, e^{- \lrb{\Gamma^0 - \Delta\Gamma/2}  \, t} \ \ket{+}\bra{+} \ + \ 
    w_{\rm M} \,  e^{- \Gamma^0  \, t} e^{+i \Delta m  \, t} \  \ket{+}\bra{-}  
       \label{eq:rho_mass_basis}\nonumber\\ 
    &+ w_- \, e^{- \lrb{\Gamma^0 + \Delta\Gamma/2}  \, t} \ \ket{-}\bra{-} \ + \
    w_{\rm M}^* \,  e^{- \Gamma^0  \, t}  e^{-i \Delta m  \, t} \  \ket{-}\bra{+}\;,
\end{align}
with 
\begin{align}
    w_+ &= |p|^2 w_{0} + |q|^2 w_{\bar 0} + p q^* w_I + q p ^* w_I^*\,, \nonumber\\
    w_- &= |p|^2 w_{0} + |q|^2 w_{\bar 0} - p q^* w_I - q p ^* w_I^*\,, \nonumber\\
    w_{\rm M} &= |p|^2 w_{0} - |q|^2 w_{\bar 0} - p q^* w_I + q p ^* w_I \;.
    \label{eq:weights_mass_basis}
\end{align}
The normalised density matrix $\hat{\rho}$ can be evaluated using~\eqs{eq:NormDenMat}, which in turn can be written in terms of the co-decaying Bloch vector $\vecb (t)$, through the relation:
$\vecb=\text{Tr}\,\big(\vecPauli \hat\rho\big)$.

Assuming that the long-lived state is $\ket{+}$, this asymptotic feature can be obtained from the density matrix $\rho_* = \rho (\infty )$, given in \eqs{eq:rho_mass_basis}, normalised to its trace, i.e.
\begin{equation}
    \hat \rho_{\star}\: =\: \dfrac{\rho_{\star}}{\text{Tr}\,\rho_{\star}}\:  =\:   \dfrac{\ket{+}\bra{+}}{\bra{+}\ket{+}}  \;.
    \label{eq:rho_inf_norm}
\end{equation}
Obviously, the system tends towards a pure state, satisfying the property:~$\hat \rho_{\star}^2 =\hat \rho_{\star}$.

In Table~\ref{tab:MesonData}, we give values for $r$, $\theta_{\mathbf{e}\boldsymbol{\gamma}}$ and $|{\bf E}|$, which reproduce the experimental central value for $\Delta m$, as well as $\Delta \Gamma$ and $\left| q/p\right|$ within the $1\,\sigma$ level of their respective central values. We note, however, that the errors on these values are highly correlated, restricting which values may be used to study critical scenarios. In particular for the $B_d^0$ system, when ${\theta = -90^\circ}$, $r$ lies close to $2 \times 10^{-3}$ limiting the reach of current experiments.

\subsection{Decoherence}

\begin{figure}[t!]
    \centering
    \includegraphics[width=0.8\linewidth]{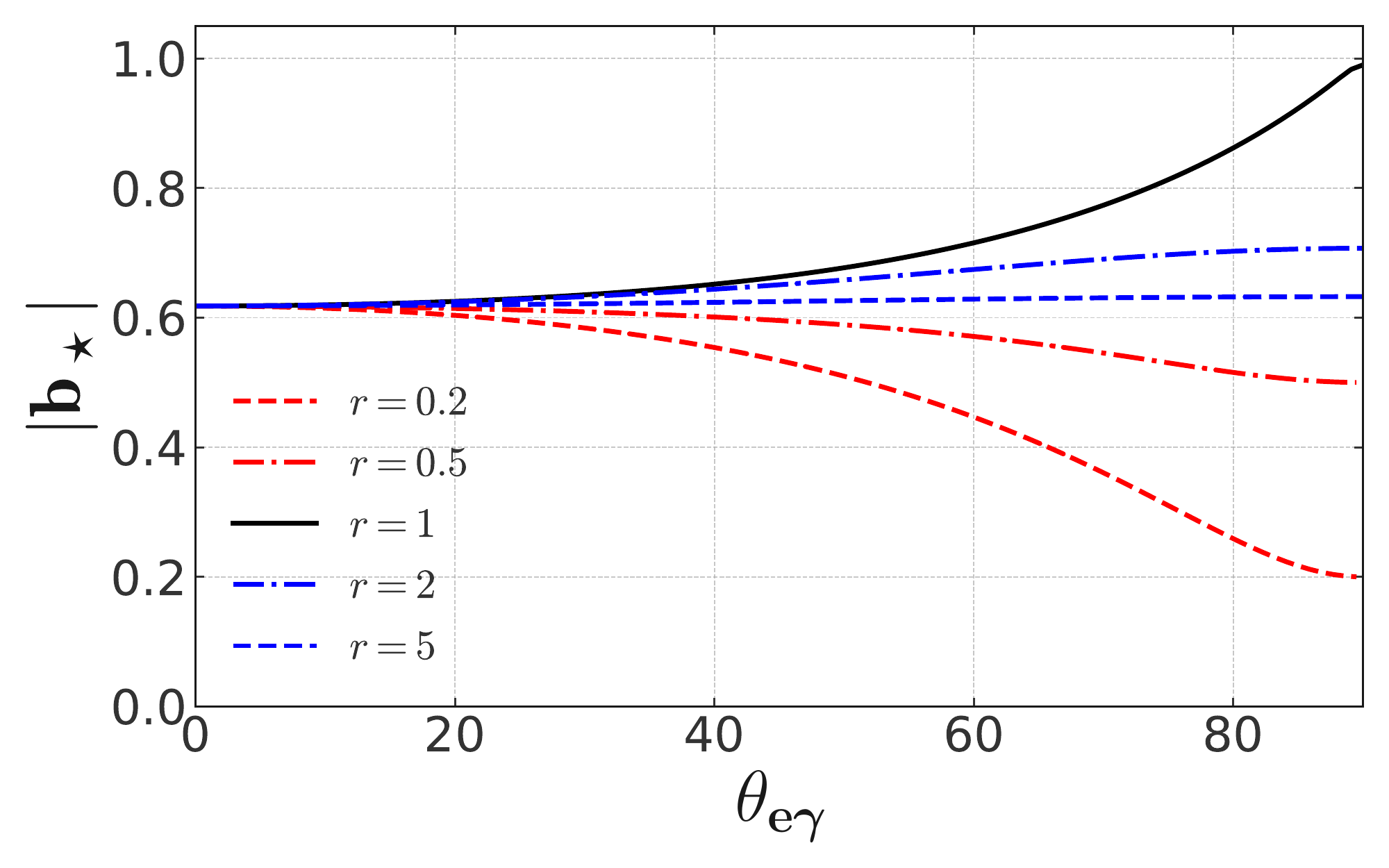}
    \caption{The magnitude of the co-decaying Bloch vector after the inclusion of decoherence effects as a function of the angle between $\mathbf{e}$ and $\boldsymbol{\gamma}$. For these results, we take $\zeta=1$, $\mathbf{b}(0) = \mathbf{e}$ and $\mathbf{d}$ perpendicular to $\mathbf{e}$ and $\boldsymbol{\gamma}$.}
    \label{fig:DecMagPlot}
\end{figure}
We expect the interaction between the system with the environment to cause decoherence. The simplest way to model such behaviour would be to add an additional term to the equation of motion for $\rho$,
\begin{equation}
    \frac{d\rho}{dt}\: =\: -i \left[\textrm{E}, \rho \right]\, -\, \frac{1}{2}\left\{\Gamma, \rho \right\}\, -\, \left[\textrm{D}, \left[\textrm{D},\rho \right] \right] \, ,
    \label{eq:rho_evolution_decoh}
\end{equation}
in close analogy to the Lindblad equation of motion \cite{cmp/1103899849, Bertlmann:2006fn}. It is clear that by taking the trace of this expression, the inclusion of this decoherence term, $\textrm{D}$, does not directly change how the trace evolves, and thus the normalisation of $\rho$ remains unaffected. However, the equation of motion for $\veca$ picks up an additional term, i.e.
\begin{equation}
    \frac{d\veca}{dt}\: =\: -2\, \mathbf{E}\times \veca\, +\, \boldsymbol{\Gamma}a^0\, -\, \Gamma^0 \veca\, +\, 4\, \mathbf{D}\times(\mathbf{D}\times\veca)\,.
\end{equation}
In this case, the equation of motion for the co-decaying Bloch vector takes on the form,
\begin{equation}
   \label{eq:dbdtdec}
    \frac{d\mathbf{b}}{d\tau}\ =\ -\,\frac{1}{r}\,(\mathbf{e}\times\mathbf{b})\: +\: \boldsymbol{\gamma}\: -\: (\boldsymbol{\gamma}\cdot\mathbf{b})\,\mathbf{b}\: +\: \zeta\, (\mathbf{d}\cdot\mathbf{b})\,\mathbf{d}\: -\: \zeta\,\mathbf{b}\,,
\end{equation}
where we have defined $\zeta =4|\mathbf{D}|^2/|\boldsymbol{\Gamma}|$, $\mathbf{d} = \mathbf{D}/|\mathbf{D}|$, and $\tau$ and $r$ were defined before in (\ref{eq:Params}). It~may then be verified that the magnitude of the co-decaying Bloch vector in the presence of decoherence phenomena will be governed by the equation of motion
\begin{equation}
   \label{eq:db2zeta}
    \frac{d|\mathbf{b}|^2}{d\tau}\: =\: 2\big(1 - |\mathbf{b}|^2 \big)\,|\mathbf{b}|\cos\theta_{\mathbf{b}\boldsymbol{\gamma}}\, -\, 2\zeta|\mathbf{b}|^2 \sin^2\theta_{\mathbf{b}\mathbf{d}}\,.
\end{equation}
With the aid of~\eqref{eq:db2zeta}, one can show that the magnitude of the co-decaying Bloch vector is stationary when
\begin{equation}
   \label{eq:bstarzeta}
    |\mathbf{b}_\star|\, =\, - \frac{\zeta}{2}\frac{\sin^2\theta_{\mathbf{b}_\star\mathbf{d}}}{\cos\theta_{\mathbf{b}_\star\boldsymbol{\gamma}}} + \sqrt{\frac{\zeta^2}{4}\frac{\sin^4\theta_{\mathbf{b}_\star\mathbf{d}}}{\cos^2\theta_{\mathbf{b}_\star\boldsymbol{\gamma}}} + 1}\;.
\end{equation}
Here, $\theta_{\mathbf{b}_\star\boldsymbol{d}}=\angle (\mathbf{b}_\star\,, \mathbf{d})$ is the angle between $\mathbf{b}_\star$ and $\mathbf{d}$. From~\eqref{eq:bstarzeta}, we recover the asymptotic result $|\mathbf{b}_\star|=1$ in the limit $\zeta\to 0$. We note that for the exceptional case with  ${\sin\theta_{\mathbf{b}_\star\boldsymbol{d}} = 0}$, we~also observe a pure asymptotic system, even in the presence of decoherence effects sourced from~${\zeta \neq 0}$.  Otherwise, unless $\mathbf{b}_\star$ is perpendicular to $\boldsymbol{\gamma}$, the inclusion of decoherence effects do not lead, in general, to fully mixed states (with $\mathbf{b} ={\bf 0}$),  since the function $f(x) = -x + \sqrt{x^2+1}$ [with $x = \zeta\, \sin^2\theta_{\mathbf{b}_\star\mathbf{d}}/(2\cos\theta_{\mathbf{b}_\star\boldsymbol{\gamma}})$] that occurs on the RHS of~\eqref{eq:bstarzeta} is always positive and has no zeros for finite $x$. Hence, decoherence effects give rise to a balancing mechanism against the purifying effect caused by the long-lived state of the system. In Fig~\label{fig:DecMagPlot}, we display the effect of decoherence phenomena on the magnitude of the co-decaying Bloch vector. As expected, the asymptotic states are neither pure nor are they completely mixed, with the exception of some special circumstances.

If we allow $\zeta \ll 1$, we find that to first order in $\zeta$, decoherence effects give a deviation in the fixed point solution (\ref{eq:FPN}),
\begin{equation}
   \label{eq:bstarJ}
    \mathbf{b}_\star\ \rightarrow\ \mathbf{b}_\star +\zeta \mathbb{J}^{-1} \mathbf{b}_\star - \zeta (\mathbf{d}\cdot \mathbf{b}_\star) \mathbb{J}^{-1} \mathbf{d}\;.
\end{equation}
Here $\mathbb{J}$ symbolises the Jacobian of the system at the fixed point, given in (\ref{eq:Jacobian}) of Appendix~\ref{app:Nonlinear}. Note that we need to take the inverse of the Jacobian, $\mathbb{J}^{-1}$, in~\eqref{eq:bstarJ}. Therefore, it is crucial that $\textrm{Det}\,\mathbb{J} \neq 0$, which is enforced by requiring that $\mathbf{b}_\star$ be a stable fixed point.

The complete time-evolution of the density matrix of unstable qubits may be determined using~\eqs{eq:dbdtdec}. Such cases were discussed in~\cite{Ellis:1983jz,Ellis:1992dz,Huet:1994kr,Bernabeu:2003ym} in the context of CPT violation due to violation of QM. Our approach is a generalisation of these discussions as we describe in detail the behaviour of the Bloch vector~$\vecb$, along with its geometrical interpretation.

\subsection{Entanglement Entropy}\label{sec:Entropy}

\begin{figure}[t!]
    \centering
    \includegraphics[width=0.8\linewidth]{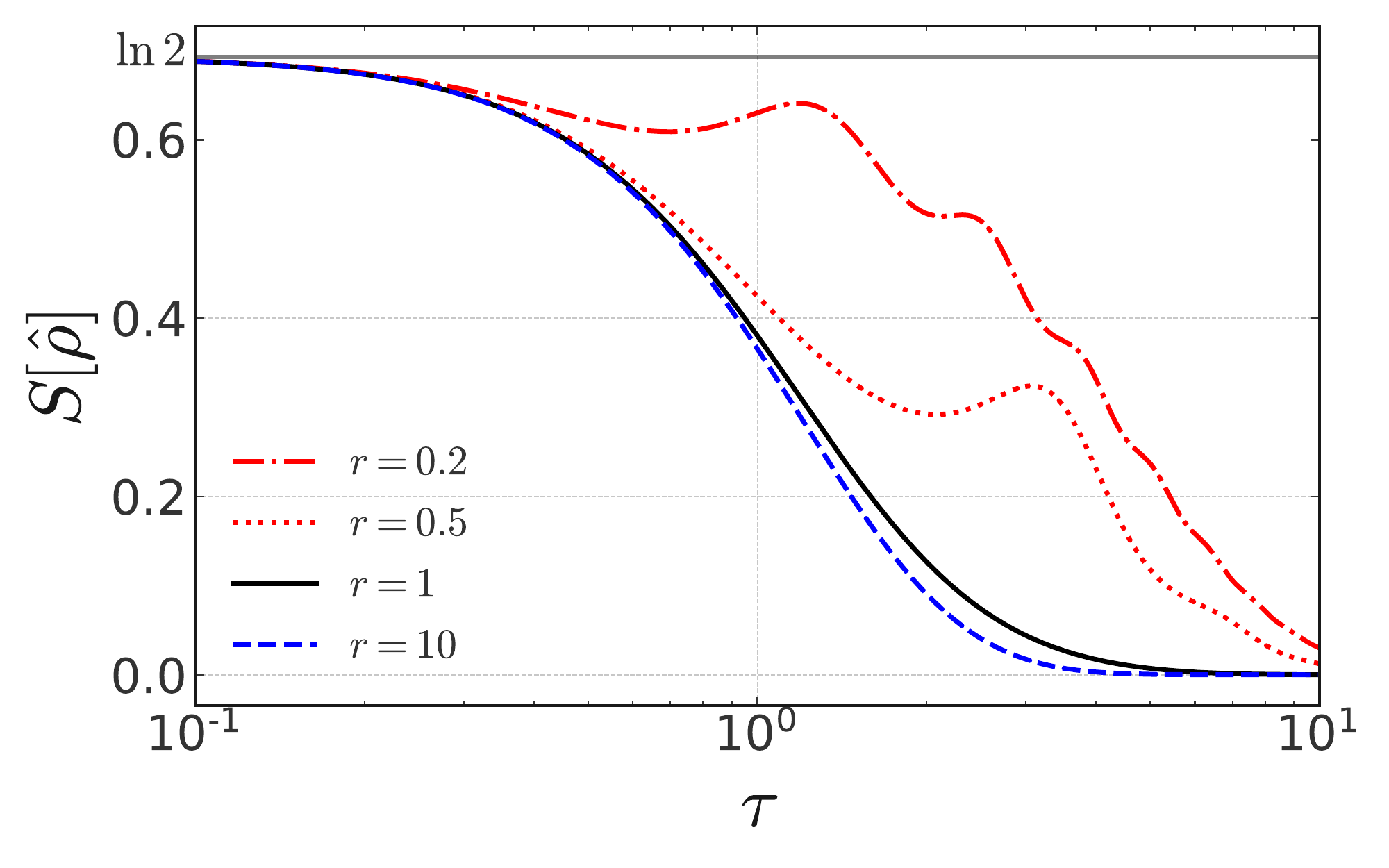}
    \caption{Entropy of the normalised system  with input parameters $\theta_{\mathbf{e}\boldsymbol{\gamma}} = 75^\circ$, and $\mathbf{b}(0)=\mathbf{0}$.}
    \label{fig:Entropy}
\end{figure}

An alternative way to analyse (de-)coherence phenomena in qubits will be to evaluate the entanglement entropy contained within the density matrix, also known as the von-Neumann entropy~\cite{vonNeumann:1955}. The entanglement entropy may be expressed as
\begin{equation}
    S[\rho]\: =\: -\,\textrm{Tr}\big(\rho\, \ln\rho \big)\,.
\end{equation}
We simply note that $S[\rho]$ satisfies identical properties to the known entropy expression from statistical mechanics and as such, it constitutes a natural extension to QM, since the diagonal elements of $\rho$ correspond to probabilities of all the different states comprising the QM system. In this sense, when $\rho$ is in its diagonal basis, the von-Neumann entropy overlaps exactly with the expression used in classical statistical mechanics. 

By rewriting the density matrix given in (\ref{eq:NormDenMat}) as $\rho =\textrm{Tr}(\rho)\,\hat{\rho} = a^0 \hat{\rho}$, one may show that the entanglement entropy of a qubit system is given by
\begin{equation}
    S[\rho]\: =\: -\,a^0\ln a^0\: +\: a^0\, S[\hat{\rho}]\: =\: -\,a^0 \ln a^0\: -\: a^0\,\textrm{Tr}\big(\hat{\rho} \ln\hat{\rho} \big)\;,
\end{equation}
neatly relating the entropy of the co-decaying system, $S[\hat{\rho}]$, to the total entropy,~$S[\rho]$. By virtue of a Taylor's series expansion of the logarithm, we may evaluate the co-decaying entropy for two-level systems in terms of the magnitude of the co-decaying Bloch vector to find
\begin{equation}\label{eq:NormEntropy}
    S[\hat{\rho}]\, =\, \ln 2 - \frac{1}{2} \left( 1 - |\mathbf{b}| \right) \ln \left( 1 - |\mathbf{b}| \right) - \frac{1}{2} \left( 1 + |\mathbf{b}| \right) \ln \left( 1 + |\mathbf{b}| \right).
\end{equation}
Observe that this last expression reproduces the known results \cite{vonNeumann:1955}, 
\begin{equation}
    S[\hat{\rho}_{\textrm{Coh}}]\, =\, 0\,, \qquad S[\hat{\rho}_{\textrm{Mix}}]\, =\, \ln 2\;,
\end{equation}
for fully coherent and fully mixed density matrices, $\hat{\rho}_{\textrm{Coh}}$ and $\hat{\rho}_{\textrm{Mix}}$, respectively, and generalises a similar expression found in \cite{deSouza:2016hrw}.
Therefore, the total entropy of the two-level system may entirely be represented in terms of the components of the Bloch vector as
\begin{equation}
    S[\rho] = - a^0\ln a^0 +  a^0\ln 2 - \frac{a^0}{2} \left( 1 - |\mathbf{b}| \right) \ln \left( 1 - |\mathbf{b}| \right) - \frac{a^0}{2} \left( 1 + |\mathbf{b}| \right) \ln \left( 1 + |\mathbf{b}| \right).
\end{equation}

In Figure~\ref{fig:Entropy}, we show the time evolution of the entanglement entropy corresponding to the density matrix in the co-decaying basis. As we would expect, the entropy of the system starts at $\ln 2$ when the two-level system is in a fully mixed state, and falls off to zero as the system purifies. For large values of $r$, the entropy of the two-level system monotonically falls to zero. However, for small values of $r$, we begin to see novel features due to (quasi-)critical scenarios. As~can be seen in Figure~\ref{fig:Entropy} for the unstable qubits with $r=0.2$ and $0.5$, the entanglement entropy of these qubits no longer exhibits a simple downward trend, but has oscillatory features as well. These features signify the onset of coherence-decoherence oscillations of the co-decaying density matrix. We will analyse these phenomena in more detail in the following sections.

\vfill\eject

\section{Critical Scenarios}\label{sec:CritScen}
\setcounter{equation}{0}

As we found in the previous section, a critical unstable two-level system is characterised by the following two properties:  $\mathbf{e}\perp\boldsymbol{\gamma}$ and $r<1$. For this class of scenarios, we~showed in~\eqref{eq:CSFP} that the asymptotic solutions are unstable. This lack of solutions for critical unstable qubits suggests that such scenarios will require a more attentive treatment. In this section, we will study this particular class of critical scenarios in more depth.

To begin with, we observe that any co-decaying Bloch vector which lies on the plane spanned by $\boldsymbol{\gamma}$ and $\mathbf{e}\times\boldsymbol{\gamma}$ will remain on this same plane at all times. This becomes apparent when we consider infinitesimal $\tau$-steps. To first order in the step difference, $\varepsilon$, we see that
\begin{align}
   \label{eq:btaueps}
    \mathbf{b}(\tau + \varepsilon) &=\ \mathbf{b}(\tau) + \varepsilon \vecb^\prime + \mathcal{O}(\varepsilon^2)\nonumber\\
    &=\ \mathbf{b}(\tau) - \frac{\varepsilon}{r}(\mathbf{e}\times\mathbf{b}(\tau)) + \varepsilon\boldsymbol{\gamma} - \varepsilon(\boldsymbol{\gamma} \cdot \mathbf{b}(\tau))\mathbf{b}(\tau) + \mathcal{O}(\varepsilon^2)\nonumber\\
    &=\ \left[ 1 - \varepsilon(\boldsymbol{\gamma}\cdot \mathbf{b}(\tau) ) \right] \mathbf{b}(\tau) + \varepsilon\boldsymbol{\gamma} - \frac{\varepsilon}{r}(\mathbf{e}\times\mathbf{b}(\tau)) + \mathcal{O}(\varepsilon^2)\;,
\end{align}
where a prime (${}^\prime$) on $\vecb (\tau )$ indicates differentiation with respect to the dimensionless parameter $\tau=|\mathbf{\Gamma}|t$.
From the last equality in~\eqref{eq:btaueps}, we see that the first two terms give rise to
a $\vecb (\tau +\varepsilon )$ which is on the $\left( \boldsymbol{\gamma}, \mathbf{e}\times\boldsymbol{\gamma} \right)$-plane, provided $\mathbf{b}(\tau)$ was on this plane. The third term also drives $\vecb (\tau +\varepsilon )$ to lie on this plane,
because of the property of the critical scenario, $\mathbf{e}\perp \boldsymbol{\gamma}$. Hence, $\mathbf{e}\times\mathbf{b}(\tau)$ will stay co-planar to $\mathbf{b}(\tau)$ at all times. Here, we must emphasize that this is a  distinct property of the critical scenario under study, since the torsion~${\cal T}$ as defined in the Frenet-Serret frame \cite{doCarmo:1976} vanishes, \ie
\begin{equation}
   \label{eq:Torsion}
    \mathcal{T}\ \equiv\ \frac{(\mathbf{b}^\prime \times \mathbf{b}^{\prime\prime})\cdot\mathbf{b}^{\prime\prime\prime}}{| \mathbf{b}^\prime \times \mathbf{b}^{\prime\prime} |^2}\ =\ 0,
\end{equation}
regardless of the choice of $\mathbf{b}(0)$ or $r = |\mathbf{\Gamma}|/2|\mathbf{E}|$, provided $\mathbf{e}$ is perpendicular to $\boldsymbol{\gamma}$.
 Consequently, trajectories of the Bloch vector $\vecb (\tau )$ are constrained to lie on a plane for a critical unstable qubit system. 

For simplicity, we consider the circumstance already discussed at the beginning of this section, with $\mathbf{b}$ constrained to the $\left( \boldsymbol{\gamma}, \mathbf{e}\times\boldsymbol{\gamma} \right)$ plane. Since we are already aware that the system will feature oscillatory effects, but will maintain its magnitude if $|\mathbf{b}(0)| = 1$, we expect to find closed loops on the plane. It is therefore logical to represent this system in cylindrical co-ordinates, with basis vectors $(\hat{\mathbf{R}}, \hat{\boldsymbol{\varphi}}, \hat{\mathbf{z}})$, and the plane co-ordinates $\left(|\mathbf{b}|, \varphi\right)$. We perform the transformation using
\begin{equation}
    \boldsymbol{\gamma} = \cos\varphi \; \hat{\mathbf{R}} - \sin\varphi \; \hat{\boldsymbol{\varphi}}, \qquad \mathbf{e}\times\boldsymbol{\gamma} = \sin\varphi \; \hat{\mathbf{R}} + \cos\varphi \; \hat{\boldsymbol{\varphi}}, \qquad \mathbf{e} = \hat{\mathbf{z}}.
\end{equation}
This representation allows the co-decaying Bloch vector to be written as $\mathbf{b} = |\mathbf{b}|\, \hat{\mathbf{R}}$, and the equation of motion may be written in terms of radial and angular components,
\begin{subequations}\label{eq:PolEoMs}
\begin{equation}\label{eq:PolEoM1}
    \frac{d|\mathbf{b}|}{d\tau}\ =\ \left( 1 - |\mathbf{b}|^2 \right)\, \cos\varphi,
\end{equation}
\begin{equation}\label{eq:PolEoM2}
    \frac{d\varphi}{d\tau}\ =\ -\,\frac{1}{r}\: -\: \frac{\sin\varphi}{|\mathbf{b}|}\;.
\end{equation}
\end{subequations}
From the set of differential equations~\eqref{eq:PolEoM1} and~\eqref{eq:PolEoM2}, a number of properties may be deduced. More explicitly, one may verify that there exists a fixed point when $r\geq 1$, and that this fixed point corresponds to a pure state $|\mathbf{b}| = 1$. But, when $r<1$, we find that there are no values for $|\mathbf{b}|$ and $\varphi$, such that the equations of motion vanish. Assuming for simplicity $r \ll |\mathbf{b}|$, then it may be shown that the magnitude of the co-decaying Bloch vector has the parametric form
\begin{equation}
    |\mathbf{b}(\tau)|\: =\: \sqrt{1\,-\,\mathcal{A}\, e^{2r\sin\varphi(\tau)}}\;,
    \label{eq:b_paramteric}
\end{equation}
where $\mathcal{A}$ is a positively valued constant determined by the initial conditions of the system. Again we notice that if we set $|\mathbf{b}|=1$, as an initial condition, the magnitude of $\mathbf{b}$ remains fixed to unity, independently of the value of $\varphi$. Furthermore, since we require that $r \ll |\mathbf{b}|$, this leads to $\mathcal{A}\ll 1$, which allows us to carry out the approximation,
\begin{equation}
   \label{eq:btau}
    |\mathbf{b}(\tau)|\: \simeq\: 1\, -\, \frac{1}{2}\mathcal{A}\, -\, \mathcal{A}r\sin \varphi (\tau)\;.
\end{equation}
Here, we should remark that $d\varphi/d\tau < 0$ implies a co-decaying Bloch vector~$\mathbf{b}(\tau)$ traversing in the clockwise direction.

\begin{figure}[t!]
    \centering
    \begin{subfigure}[t]{0.49\linewidth}
    \centering
    \includegraphics[width=\linewidth]{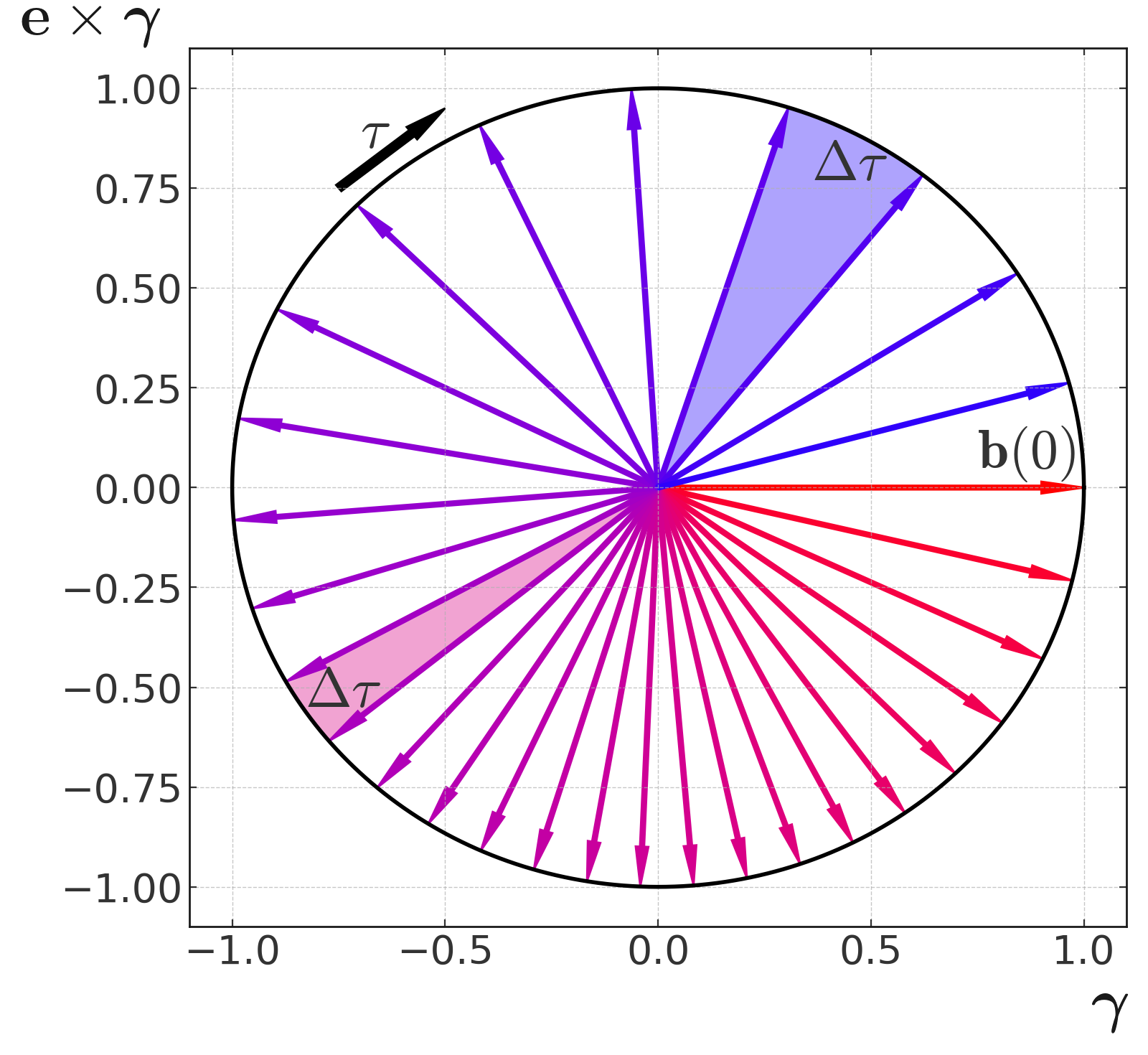}
    \caption{}
    \label{fig:CSBlochP}
    \end{subfigure}
    \begin{subfigure}[t]{0.49\linewidth}
    \centering
    \includegraphics[width=\linewidth]{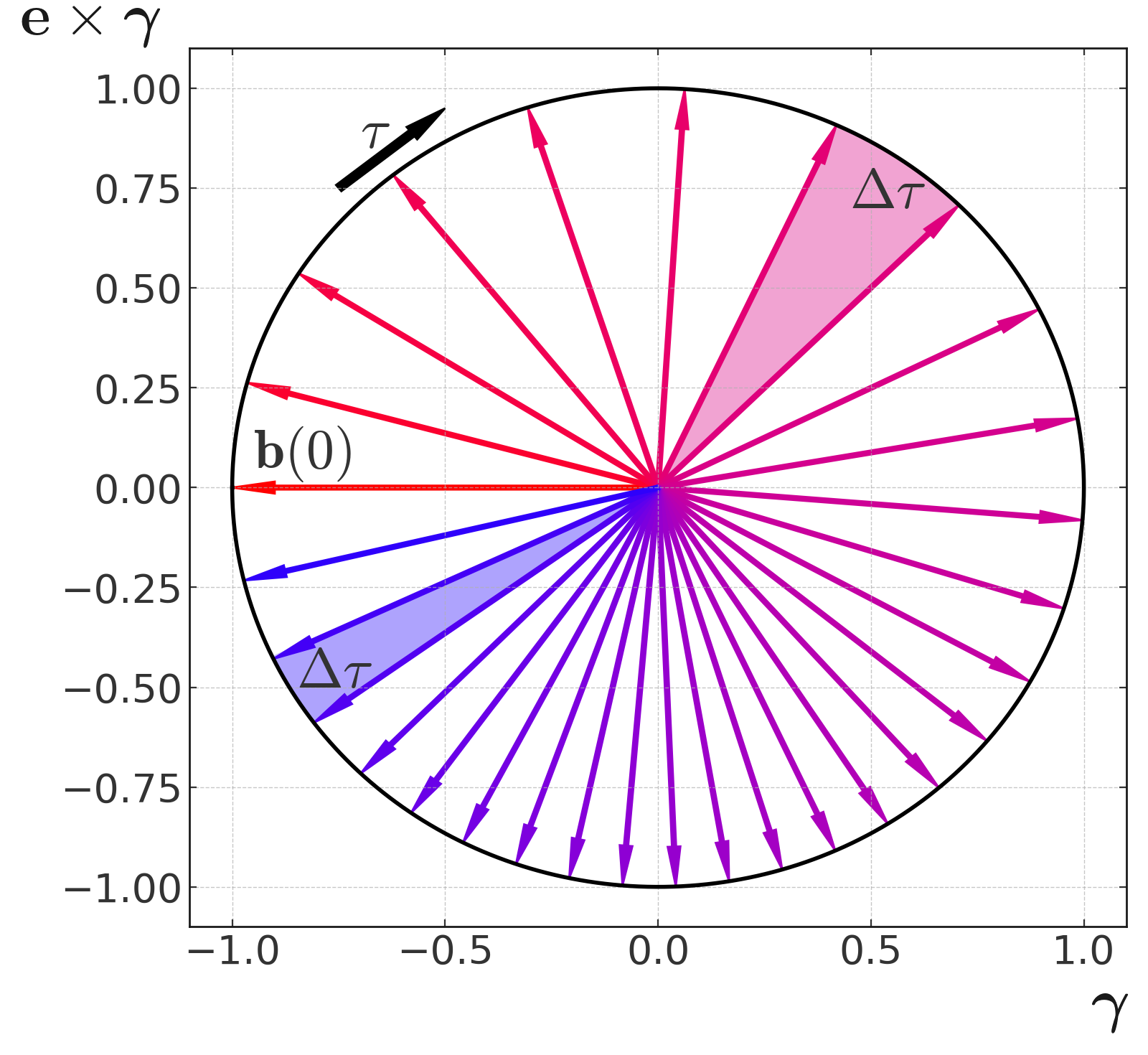}
    \caption{}
    \label{fig:CSBlochN}
    \end{subfigure}
    \caption{Evolution of the co-decaying Bloch vector ${\bf b}(\tau )$ in a critical scenario with $r=0.5$, for: (a)~$\mathbf{b}(0) = \boldsymbol{\gamma}$ and (b) $\mathbf{b}(0) = -\boldsymbol{\gamma}$. In a colour varying manner, red arrows represent the co-decaying Bloch vector close to~$\tau = 0$, whilst the co-decaying Bloch vector approaching~$\tau = \widehat{\textrm{P}}$ is given in blue.}
    \label{fig:CSBloch}
\end{figure}

Substituting now~\eqref{eq:btau}  into~\eqref{eq:PolEoM2} and integrating over the full range of $\varphi\in [2\pi , 0]$ gives an approximate expression for the dimensionless (Lorentz-invariant) period of oscillation, denoted here as $\widehat{\textrm{P}}$, for the critical scenario, i.e.
\begin{equation}
   \label{eq:Phat}
    \widehat{\textrm{P}}\: \simeq\: \frac{2\pi r}{1- \mathcal{A}} \, \Bigg\{\bigg[1-\left(\frac{2-2\mathcal{A}}{2-\mathcal{A}} \right)^2 r^2\bigg]^{-1/2} -\, \mathcal{A}\Bigg\}\;.
\end{equation}
For $\mathcal{A}=0$, the two approximate expressions, for 
$|\mathbf{b}(\tau)|$ and $\widehat{\textrm{P}}$ in~\eqref{eq:btau} and~\eqref{eq:Phat}, assume a simple and exact analytic form,
 \begin{equation}
     |\mathbf{b}|\, =\, 1\,, \qquad \widehat{\textrm{P}}\, =\, \frac{2\pi r}{\sqrt{1-r^2}}\; .
     \label{eq:A=0_crit}
 \end{equation}
Restoring the time dimension for the period $\text{P}$ of oscillation in natural units yields
\begin{equation}
   \label{eq:Posc}
    \textrm{P}\, =\, \widehat{\textrm{P}}\,|\boldsymbol{\Gamma}|^{-1} \, =\, \frac{\pi}{|\mathbf{E}|\sqrt{1-r^2}}\; .
\end{equation}
Note that for $r=0$, \eqref{eq:Posc} gives the well-known expression for the time period of Rabi or qubit oscillations.

The fact that $|\mathbf{b}|$ has oscillatory behaviour implies that in the critical scenario, it is possible for the two-level system to transition between near-coherent and mixed states in a predictable manner. From this result, and the expression given in (\ref{eq:NormEntropy}), we see that these transitions between mixed and near-coherent states is the origin of the non-trivial behaviour for the entropy of the system shown in Figure~\ref{fig:Entropy}.
 
Moreover, the dependence of (\ref{eq:PolEoM2}) on $\sin \varphi $ leads to inhomogeneity within the angular speed of the co-decaying Bloch vector. As can be seen in (\ref{eq:PolEoM2}), the angular speed is of greatest magnitude when $\sin \varphi  > 0$, and hence we expect to see the co-decaying Bloch vector rotate faster in this region. As a result, we expect that the co-decaying Bloch vector will sweep out unequal areas in equal times.

In Figure~\ref{fig:CSBloch}, we present numerical results for the time evolution of the co-decaying Bloch vector~${\bf b}(\tau)$. As initial conditions for the critical unstable two-level systems of interest here, we set $|\mathbf{b}(0)| = 1$, corresponding to $\mathcal{A}=0$, for which the exact solution \eqs{eq:A=0_crit} applies. To create the two plots in Figure~\ref{fig:CSBloch}, we took adjacent arrows to be equally $\tau-$spaced, and therefore closer packed arrows indicate where the co-decaying Bloch 
vector~${\bf b}(\tau)$ is moving slower. As expected, the two plots exhibit a clockwise rotation of the co-decaying Bloch vector~${\bf b}(\tau)$ with a faster motion in the upper half plane. This is beautifully illustrated in Figure~\ref{fig:CSBloch} by colour-varying arrows representing the vector~${\bf b}(\tau)$.
In addition, as can be seen through the highlighted segments, the co-decaying Bloch vector sweeps out a different area in the lower half plane than it does in the upper half plane for identical $\Delta \tau$. This crucially differs from the result one would expect if we were to take the decay and oscillations as separate effects. In the case that there are no decays, the rotation of the co-decaying Bloch vector occurs at a constant rate. Thus, the proper inclusion of decay effects is of paramount importance when analysing critical unstable qubit systems.

\begin{figure}[t!]
    \centering
    \begin{subfigure}[t]{0.49\linewidth}
    \centering
    \includegraphics[width=\linewidth]{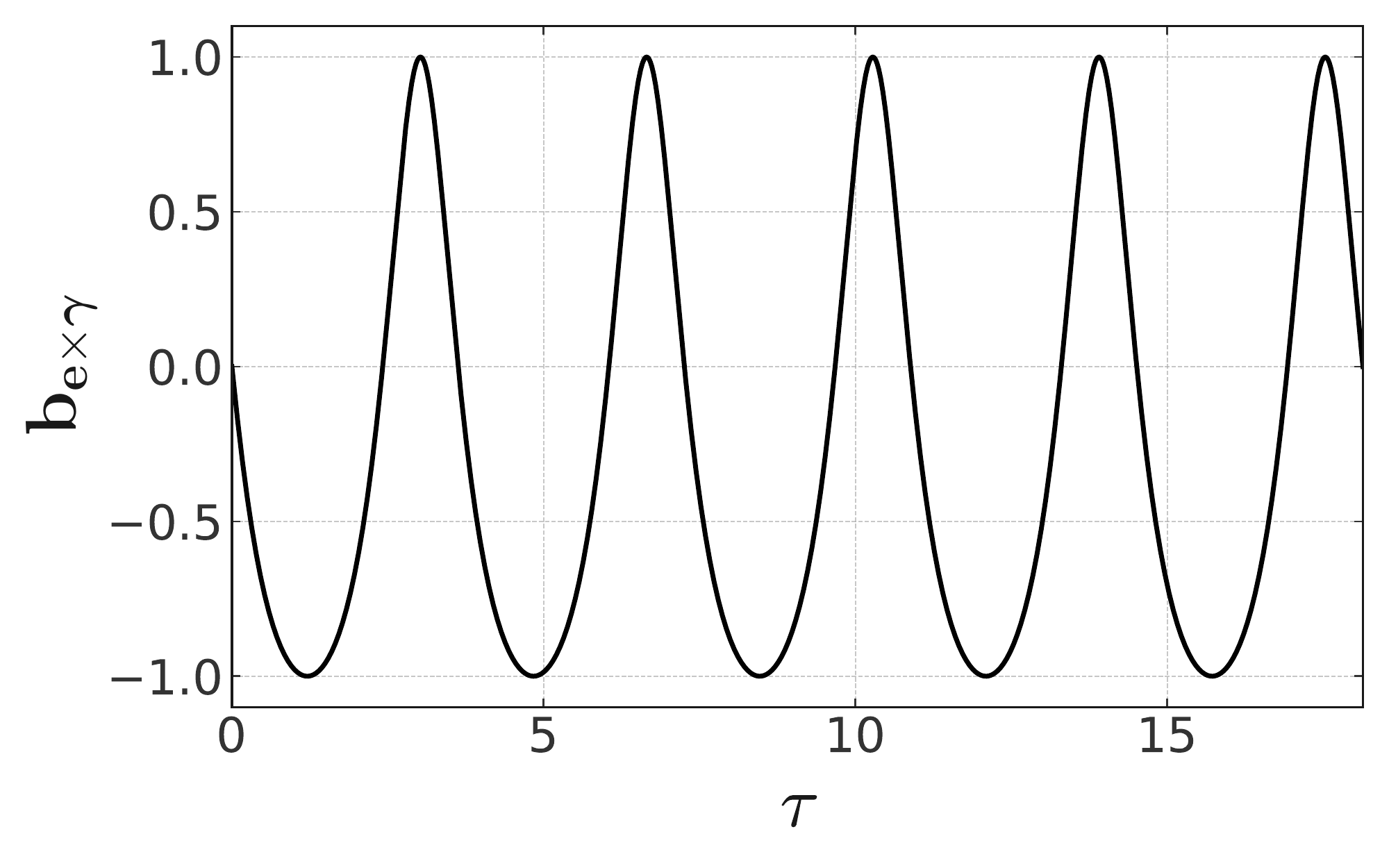}
    \caption{}
    \label{fig:EGProA}
    \end{subfigure}
    \begin{subfigure}[t]{0.49\linewidth}
    \centering
    \includegraphics[width=\linewidth]{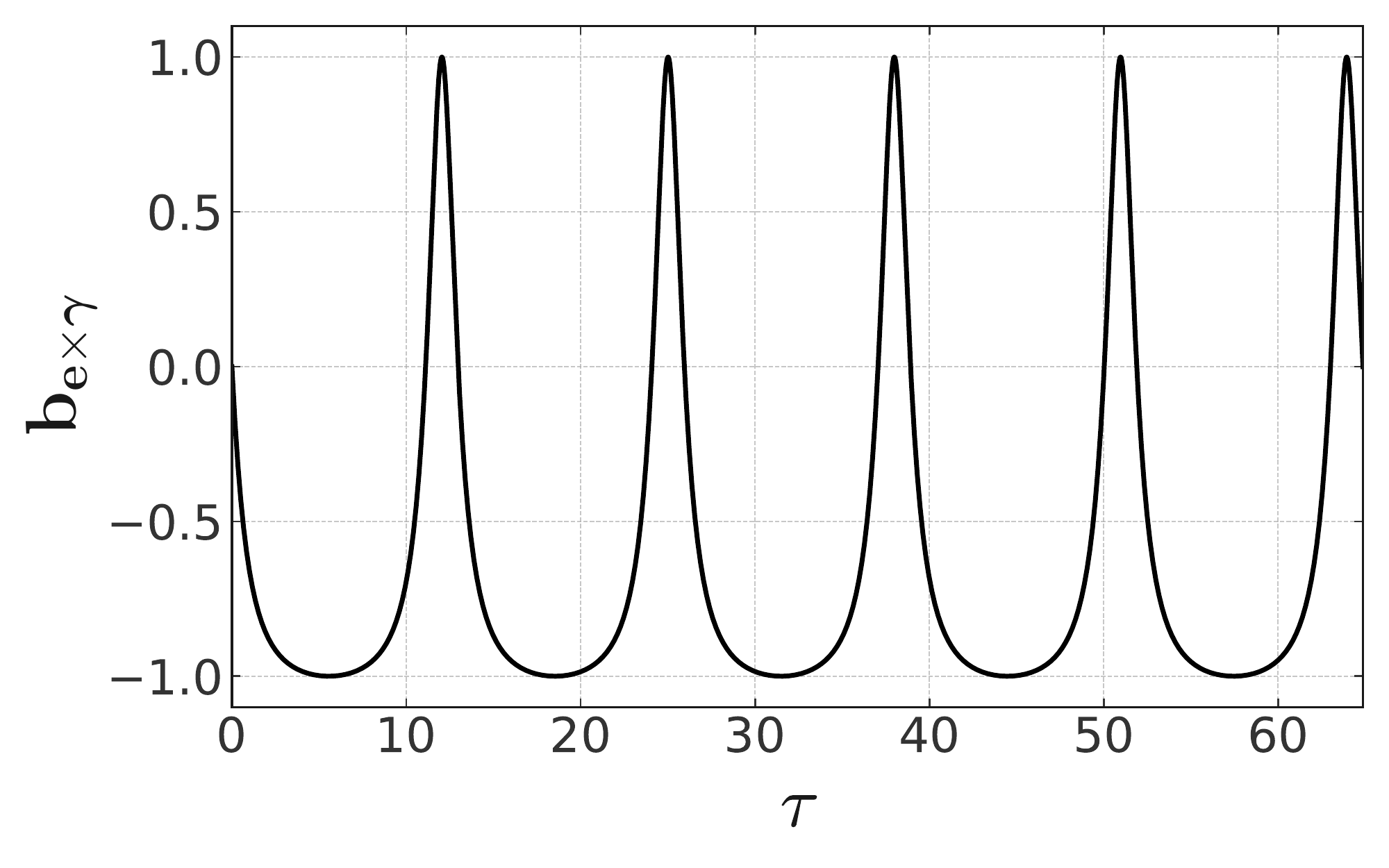}
    \caption{}
    \label{fig:EGProB}
    \end{subfigure}
    \caption{The projection of the co-decaying Bloch vector, $\mathbf{b}$, along the direction $\mathbf{e}\times \boldsymbol{\gamma}$ for: (a)~$r=0.5$ and (b)~$r=0.9$ with initial condition $\mathbf{b}(0) = \boldsymbol{\gamma}$.}
    \label{fig:EGPro}
\end{figure}

In Figure~\ref{fig:EGPro}, we show how the inhomogenous angular frequency may appear within an experimental setting. As can be seen in the two plots, if the co-decaying Bloch vector~${\bf b}(\tau)$ is projected onto a particular axis, e.g.~along ${\bf e}\times\boldsymbol{\gamma}$, the oscillations present on this axis may be distorted away from the harmonic  oscillations one may expect. For low values of $r$, oscillations are close to harmonic, as shown in Figure~\ref{fig:EGProA}. But, for scenarios with $r$ closer to $1$, we observe a large degree of distortion from their sinusoidal form, as reflected in Figure~\ref{fig:EGProB}. In particular, we observe sharper peaks and shallower troughs in the projected Bloch vector,~${\bf b}_{{\bf e}\times\boldsymbol{\gamma}}$, than one would naively expect. Consequently, one may probe the value of $r$ by looking at the degree of distortion in the oscillatory pattern of~${\bf b}_{{\bf e}\times\boldsymbol{\gamma}}$, by studying the amplitudes of higher harmonics in a Fourier series expansion.

\subsection{Coherence--Decoherence Oscillations}

As another important application of the analysis presented in this section, let us consider a critical unstable qubit which is initially prepared in a fully mixed state with a vanishing Bloch vector, i.e.~$\vecb(0) = \mathbf{0}$, at $t=\tau=0$. 
In this case, the approximation \eqs{eq:b_paramteric} can no longer be applied. However, this particular initial condition simplifies significantly the analytic form of the solution given in~\eqs{eq:a_solution} in terms of the co-decaying Bloch vector, which now reads
\begin{equation}
    |{\bf b}(\tau)|^2\ =\ 1\: -\:  (1-r^2)^2\,\lrsb{1-r^2 \, \cos \lrb{\dfrac{\sqrt{1-r^2}}{r} \,\tau} }^{-2}   \;.
    \label{eq:b_crit}
\end{equation}
Notice that for $r<1$, the magnitude of the Bloch vector, $|\vecb(\tau )|$, will oscillate between zero and a maximum value given by ${\rm max}(\absb) = 1 - (1-r^2)^2/(1+r^2)^2$, while the period of oscillation will be the same as the one stated in~\eqs{eq:A=0_crit}. 

\begin{figure}
    \centering
    \includegraphics[width=0.8\linewidth]{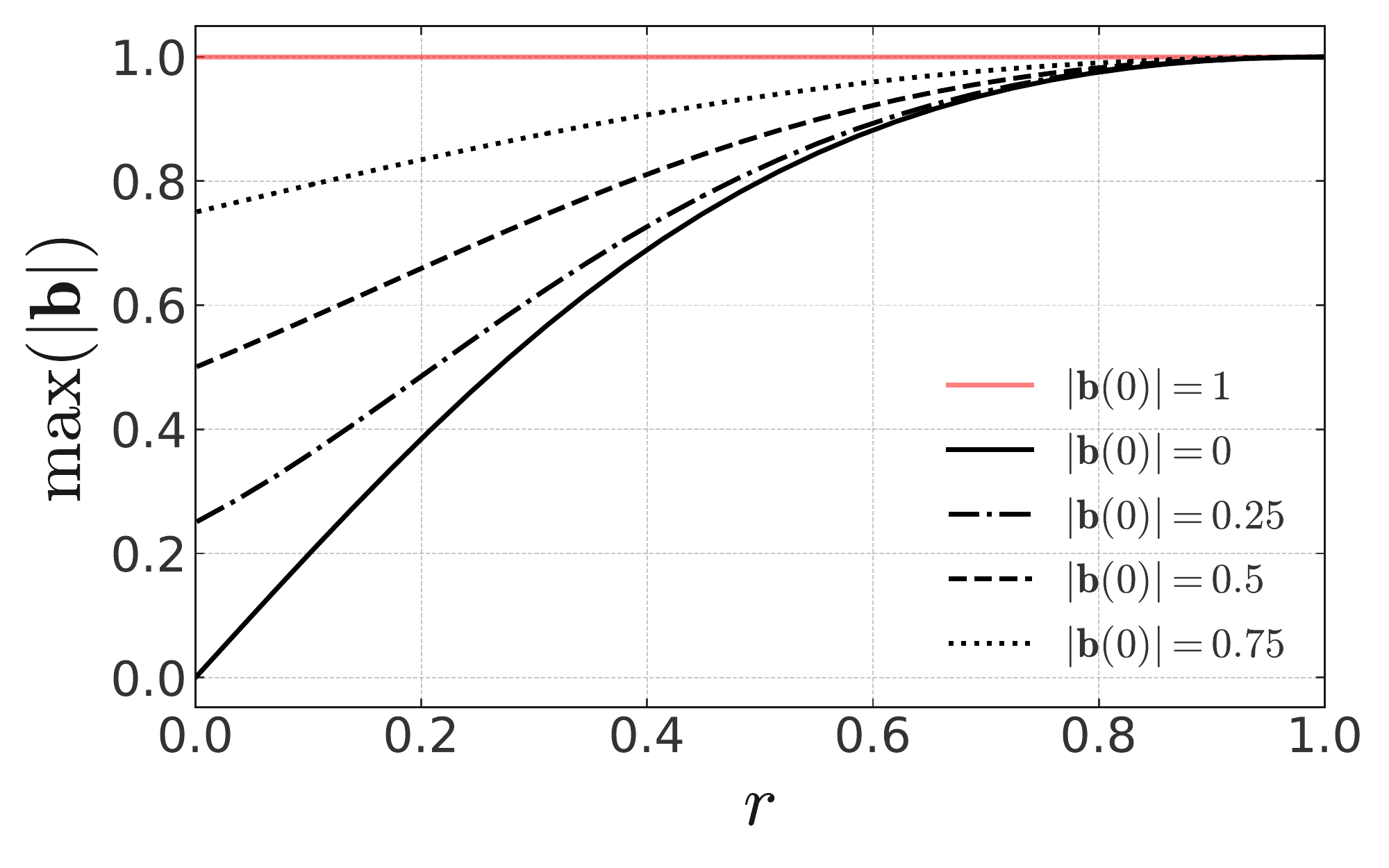}
    \caption{The maximum value of the magnitude of the co-decaying Bloch vector $\mathbf{b}(\tau )$ as a function of the parameter~$r$, for a critical unstable qubit. A set of different initial conditions for $\mathbf{b}(0)\parallel \boldsymbol{\gamma}$ was assumed.}
    \label{fig:bMax}
\end{figure}

Figure~\ref{fig:bMax} shows the maximum values of~$|\mathbf{b}(\tau )|$ as a function of~$r$, for a critical unstable qubit,
while assuming a set of representative initial conditions for~$|\mathbf{b}(0)|$
and taking $\mathbf{b}(0)\parallel \boldsymbol{\gamma}$.
Most remarkably, we find that critical unstable qubit systems will display {\em coherence--decoherence} oscillations in time when analysed in its co-decaying frame. This is in stark contrast to the dynamics of a stable qubit, where no such phenomena are possible. 

For the initial condition ${\bf b}(0) ={\bf 0}$, the aforementioned coherence--decoherence oscillatory phenomenon may intuitively be understood with the help of Figure~\ref{fig:CSBloch} as well. Specifically, a mixed Bloch vector ${\bf b}(\tau)$ may approximately be represented as a linear superposition of two {\em pure} Bloch vectors ${\bf b}_{1,2} (\tau)$ (satisfying $|{\bf b}_{1,2}(\tau)| = 1$), such that ${\bf b}(\tau) \simeq \frac{1}{2}\,\big({\bf b}_1 (\tau) + {\bf b}_2 (\tau)\big)$, with~${{\bf b}_1 (0)= \boldsymbol{\gamma} =-\,{\bf b}_2(0)}$. While the critical unstable qubit is initially in a fully mixed state (with ${\bf b}(0) = {\bf 0}$), a net non-zero Bloch vector ${\bf b}(\tau)$ will develop with time, since ${\bf b}_1 (\tau)$ and ${\bf b}_2(\tau)$ evolve differently as shown in the panels~(a) and~(b) of Figure~\ref{fig:CSBloch}, respectively.
The maxima and minima of this coherence--decoherence oscillating pattern resulting from a critical unstable qubit may be probed in an experiment by placing a detector at different positions and accumulating data in the predicted minima and maxima of~$\absb$. 
Finally, we should comment that in the limiting case $r = 1$, we have 
\begin{equation}
|{\bf b}(\tau)|^2\: =\: 1\,-\,\frac{4}{(2+\tau^2)^2}\;,
\end{equation}
so the critical unstable qubit will tend asymptotically to a pure state in the co-decaying frame, as~${\tau \to \infty}$.

\vfill\eject

Applying the above results to CP-violating critical unstable qubits, for which $\vecE \perp \vecG$ and $r<1$, the formulae for their key kinematic parameters presented in Section~\ref{sec:Meson} simplify to 
\begin{eqnarray}
    \Delta m &=& 2 \absE\, \sqrt{ 1-r^2 }\,,\qquad 
    \Delta \Gamma \ =\ 0\,, \label{eq:meson_params_perp}\nonumber\\[3mm]
    \left|\dfrac{q}{p}\right| &=& \left\{ \begin{array}{cc}
         \sqrt{\dfrac{1+r}{1-r}}\;, & \quad \theta_{\mathbf{e}\boldsymbol{\gamma}} = - \dfrac{\pi}{2} \\[0.5cm]
         \sqrt{\dfrac{1-r}{1+r}}\;, & \quad \theta_{\mathbf{e}\boldsymbol{\gamma}} = +\dfrac{\pi}{2}
    \end{array}\right. \;.
\end{eqnarray}  
From the magnitude of $q/p$, we may then infer the branch realised in a system. For example, the experimental value of the modulus of this ratio for the $B_{\rm d}^0$-meson system is larger than unity~\cite{HFLAV:2022pwe}. In particular, from Table~\ref{tab:MesonData}, we see that the $B^0_d$--$\bar{B}^0_d$ system can, in principle, realise a critical unstable two-level system, with $\theta_{\mathbf{e}\boldsymbol{\gamma}} = 270^\circ$ and $r \simeq 2 \times 10^{-3}$, in agreement with the current experimental uncertainties.

\subsection{Inclusion of Decoherence Effects}

\begin{figure}[t!]
    \centering
    \includegraphics[width=0.6\linewidth]{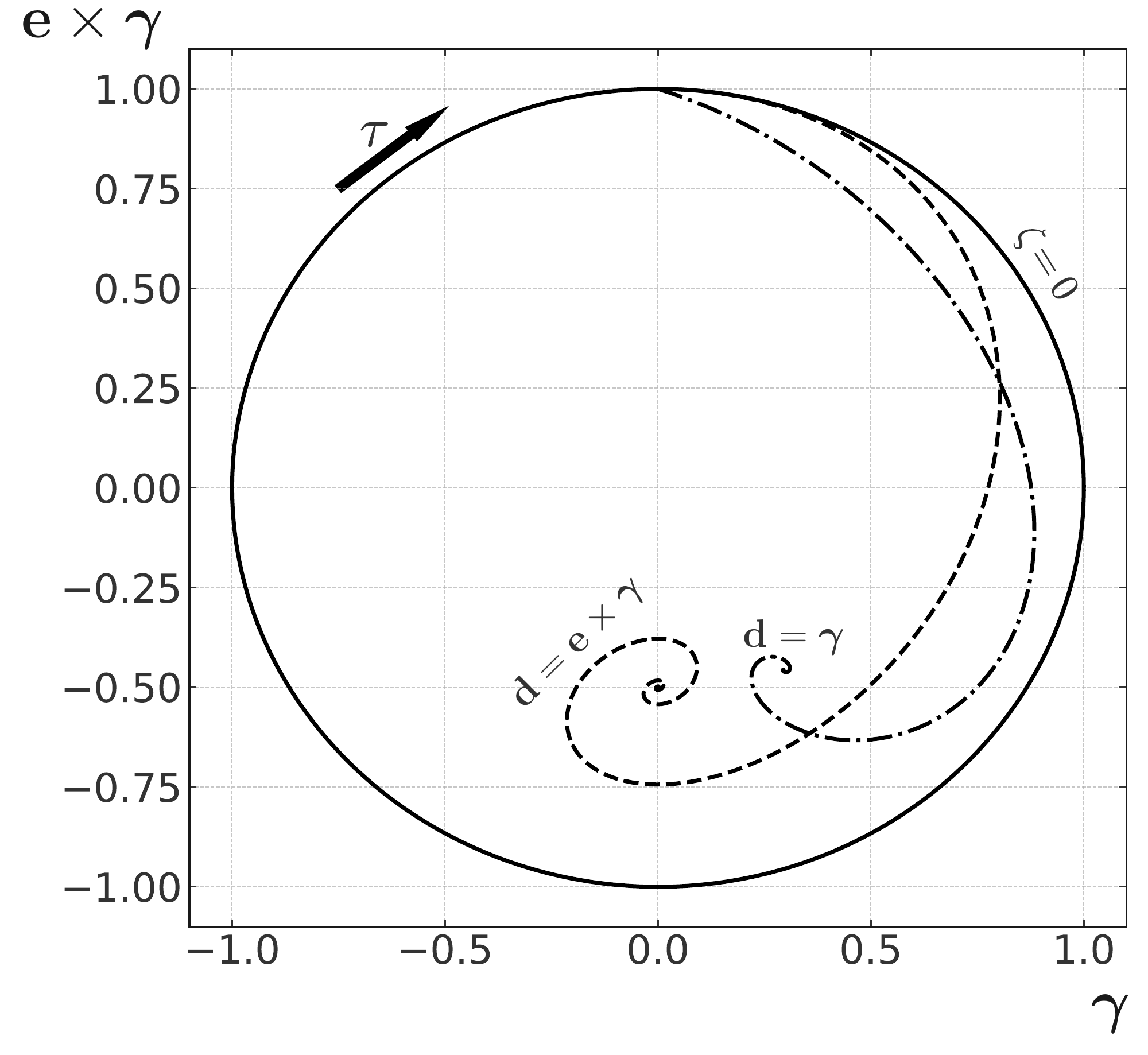}
    \caption{Trajectories for the co-decaying Bloch vector, $\mathbf{b}$, with the inclusion of decoherence effects in critical scenarios. Here, $\mathbf{d}=\boldsymbol{\gamma}$ and $\mathbf{d}=\mathbf{e}\times\boldsymbol{\gamma}$ are exhibited with $\zeta = 1$. Also included is a comparative trajectory where there are no decoherence effects, $\zeta = 0$. Results are shown for $r=0.5$ and $\mathbf{b}(0) = \mathbf{e}\times\boldsymbol{\gamma}$.}
    \label{fig:Decoherence_Fig}
\end{figure}

We will now investigate the impact of decoherence effects on critical unstable two-level systems. To this end, we will see that decoherence effects lead to the breakdown of a number of properties from the coherent system. Notably, when one calculates the torsion ${\cal T}$ defined in~\eqref{eq:Torsion}, with $\zeta \neq 0$, a non-zero result is found. Hence, the motion of the co-decaying Bloch vector ${\bf b}(\tau )$ is no longer confined to a plane, and it can spiral off. In addition, when decoherence effects are considered, self-consistent fixed point solutions in the critical scenario can be derived. For instance, when $\mathbf{d}=\mathbf{e}\times\boldsymbol{\gamma}$, we get an adjustment to the expression for $\mathbf{b}_\star$, for values 
of~${r>1}$, i.e.
\begin{equation}
    \mathbf{b}_\star = \left(\sqrt{1-\frac{1}{r^2} + \frac{\zeta^2}{4}} -\frac{\zeta}{2} \right) \boldsymbol{\gamma} - \frac{1}{r} \mathbf{e}\times \boldsymbol{\gamma}.
\end{equation}
However, we also find that for $r<1$, there exists the new asymptotic solution
\begin{equation}\label{eq:newFP}
    \mathbf{b}_\star\: =\: -\,r\, \mathbf{e}\times \boldsymbol{\gamma}\;.
\end{equation}
When we analyse the Jacobian of this system at the new fixed point given in~\eqref{eq:Jacobian}, we get a better insight why this change in behaviour occurs. Calculating the determinant and trace of the Jacobian at the point given in (\ref{eq:newFP}), we find
\begin{equation}
    \textrm{Det}\,\mathbb{J}\: =\: \zeta \left(1 - \frac{1}{r^2} \right), \qquad \textrm{Tr}\,\mathbb{J}\: =\: -2\zeta.
\end{equation}
Following the method outlined in Appendix~\ref{app:Nonlinear}, we see that when $\zeta = 0$, both $\textrm{Det}\,\mathbb{J}$ and $\textrm{Tr}\,\mathbb{J}$ vanish, implying that this point is unstable. Consequently, this gives rise to the observed oscillatory behaviour of ${\bf b}(\tau )$. However, when $\zeta\neq 0$, we see that a stable solution exists for all values of~${r<1}$. Finally, we note that for an exceptional critical scenario with~$r=1$ in which~${\mathbf{d} = \mathbf{e}\times \boldsymbol{\gamma}}$, the critical unstable qubit will eventually go to a pure state, in spite of the presence of decoherenece phenomena.

Figure~\ref{fig:Decoherence_Fig} shows the trajectory of the co-decaying Bloch vector~${\bf b}(\tau )$ with and without the inclusion of decoherence terms. As can be seen in this figure, their inclusion pulls ~${\bf b}(\tau )$ away from the oscillatory trajectory which we expect to see in the critical scenario. Instead, these trajectories now tend towards a particular point, as viewed in an appropriate slice of the Bloch sphere. In the following section, we will analyse unstable qubit systems for which
the energy-level vector ${\bf E}$ is not exactly perpendicular 
to the decay-width vector ${\bf \Gamma}$.

\section{Quasi-Critical Scenarios}\label{sec:QuasiCrit}
\setcounter{equation}{0}

To assess the significance of our results, it is important to analyse the stability of critical unstable qubits under small departures from their criticality condition: $\mathbf{e}\perp\boldsymbol{\gamma}$. 
As a consequence, it is worth considering quasi-critical scenarios, where $r<1$, but the three-vectors $\mathbf{e}$ and $\boldsymbol{\gamma}$ are not perfectly orthogonal to one another. From the analysis carried out thus far, it is clear that a generic non-critical unstable qubit will relax into a pure state as determined by the longest-lived state of the effective Hamiltonian. On the other hand, if the considered deviation away from the critical scenario is sufficiently small, it is reasonable to expect that some of the dynamics in the critical scenario will be preserved. For example, as we will see in this section, oscillations between coherent and mixed states will still be present.

\begin{figure}[t!]
    \centering
    \begin{subfigure}[t]{0.49\textwidth}
    \centering
    \includegraphics[width=\linewidth]{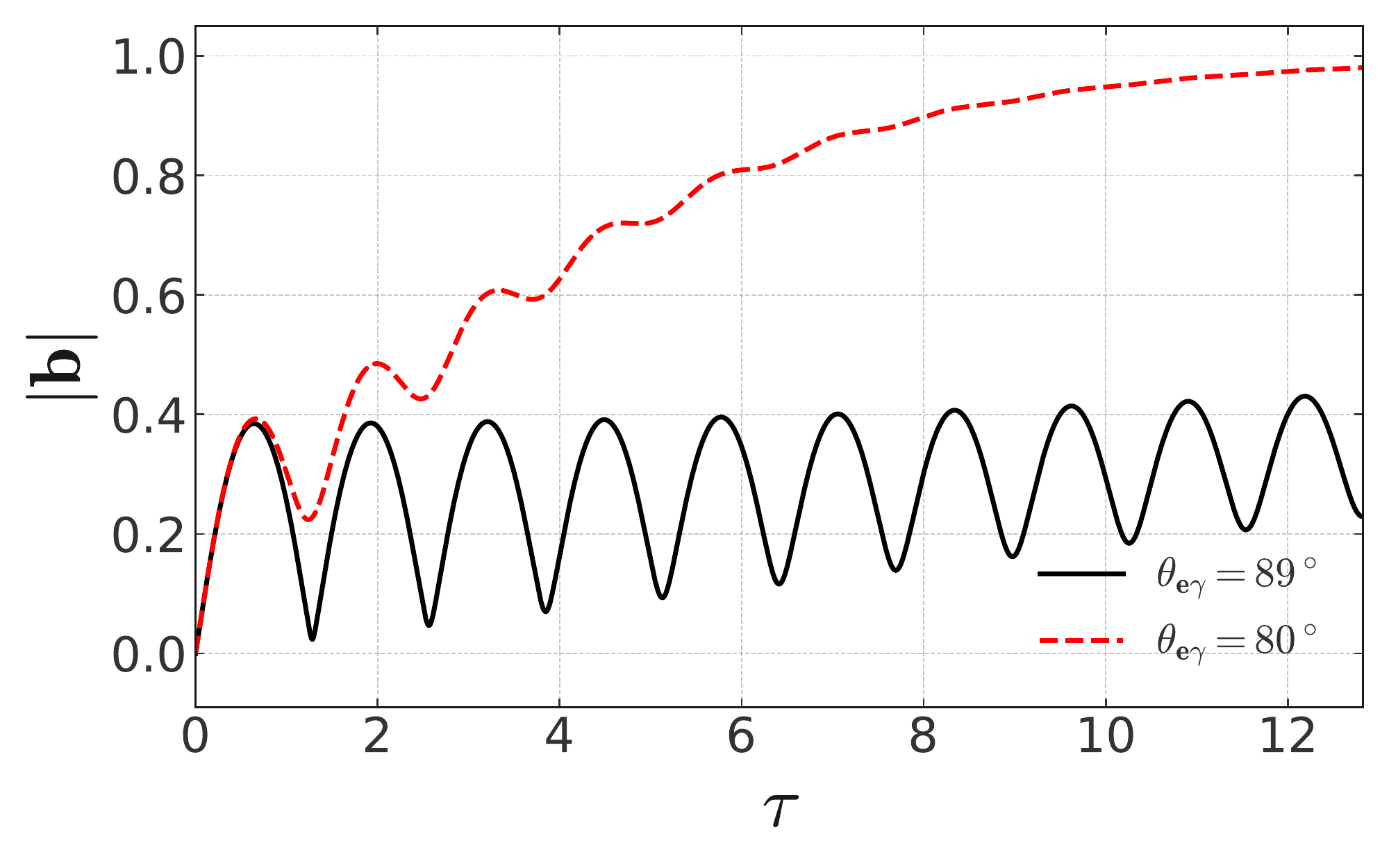}
    \caption{}
    \label{fig:ThetaEnvelopeA}
    \end{subfigure}
    \begin{subfigure}[t]{0.49\textwidth}
    \centering
    \includegraphics[width=\linewidth]{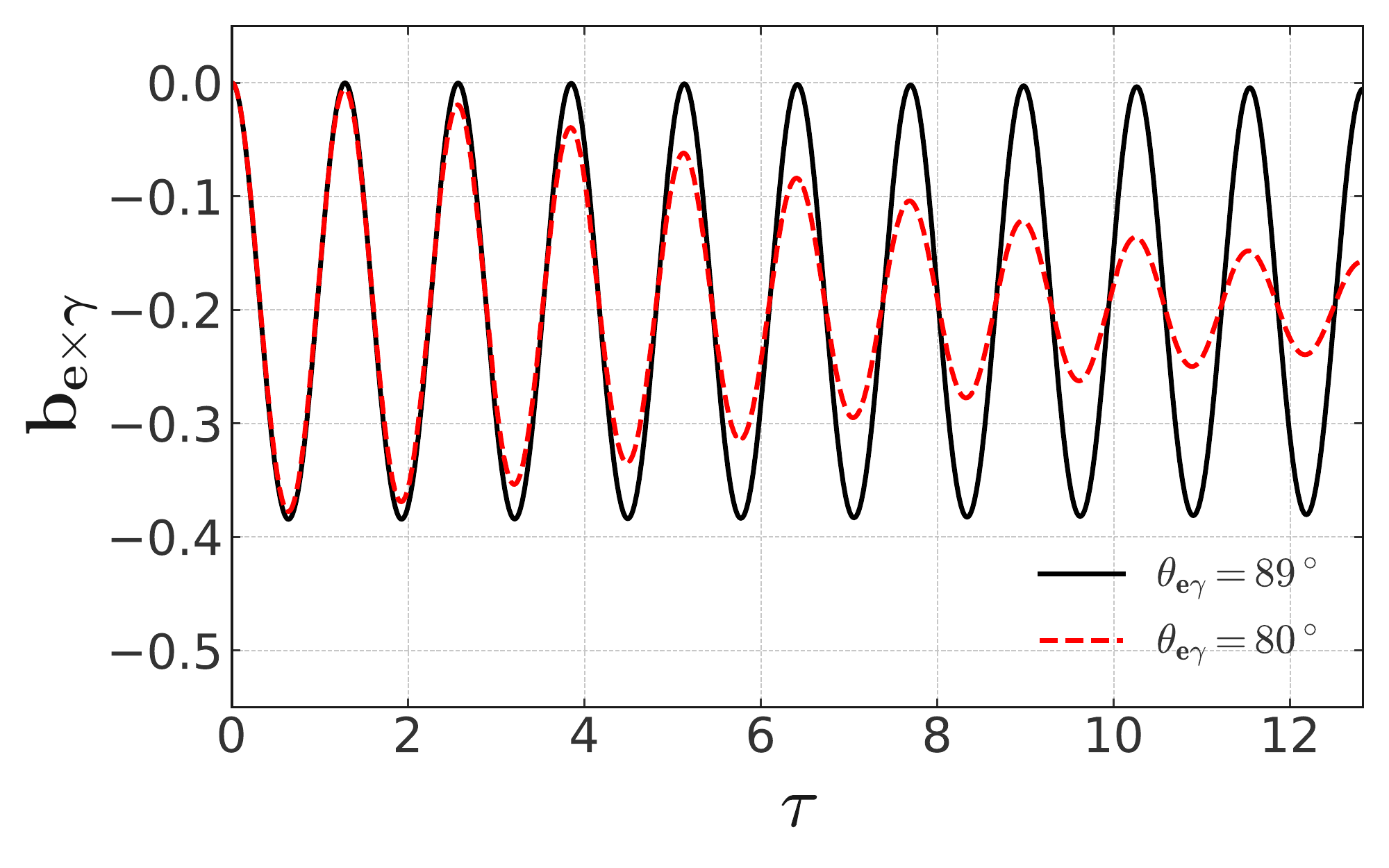}
    \caption{}
    \label{fig:ThetaEnvelopeB}
    \end{subfigure}
    \caption{ (a) The magnitude of the co-decaying Bloch vector, $\absb$. (b) Projection of the co-decaying Bloch vector in the direction $\mathbf{e} \times \boldsymbol{\gamma}$. Two quasi-critical scenarios are considered with angles $\theta_{\mathbf{e}\boldsymbol{\gamma}} = 80^\circ$ and $89^\circ$, $r=0.2$, and an initial condition of $\mathbf{b}(0) =\mathbf{0}$.}
    \label{fig:ThetaEnvelope}
\end{figure}

\begin{figure}[t!]
    \centering
    \begin{subfigure}[t]{0.49\textwidth}
    \centering
    \includegraphics[width=\linewidth]{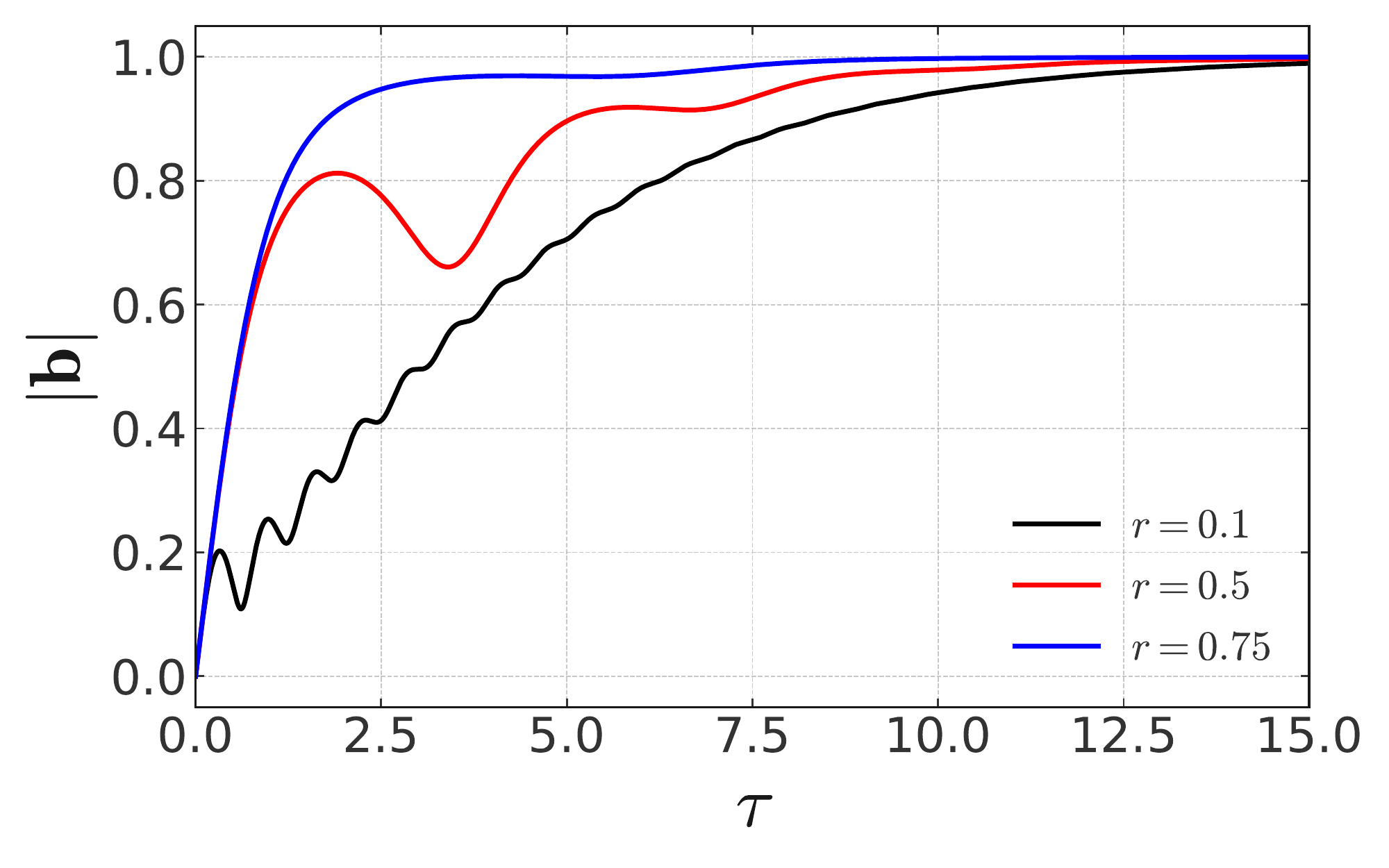}
    \caption{}
    \label{fig:rEnvelopeA}
    \end{subfigure}
    \begin{subfigure}[t]{0.49\textwidth}
    \centering
    \includegraphics[width=\linewidth]{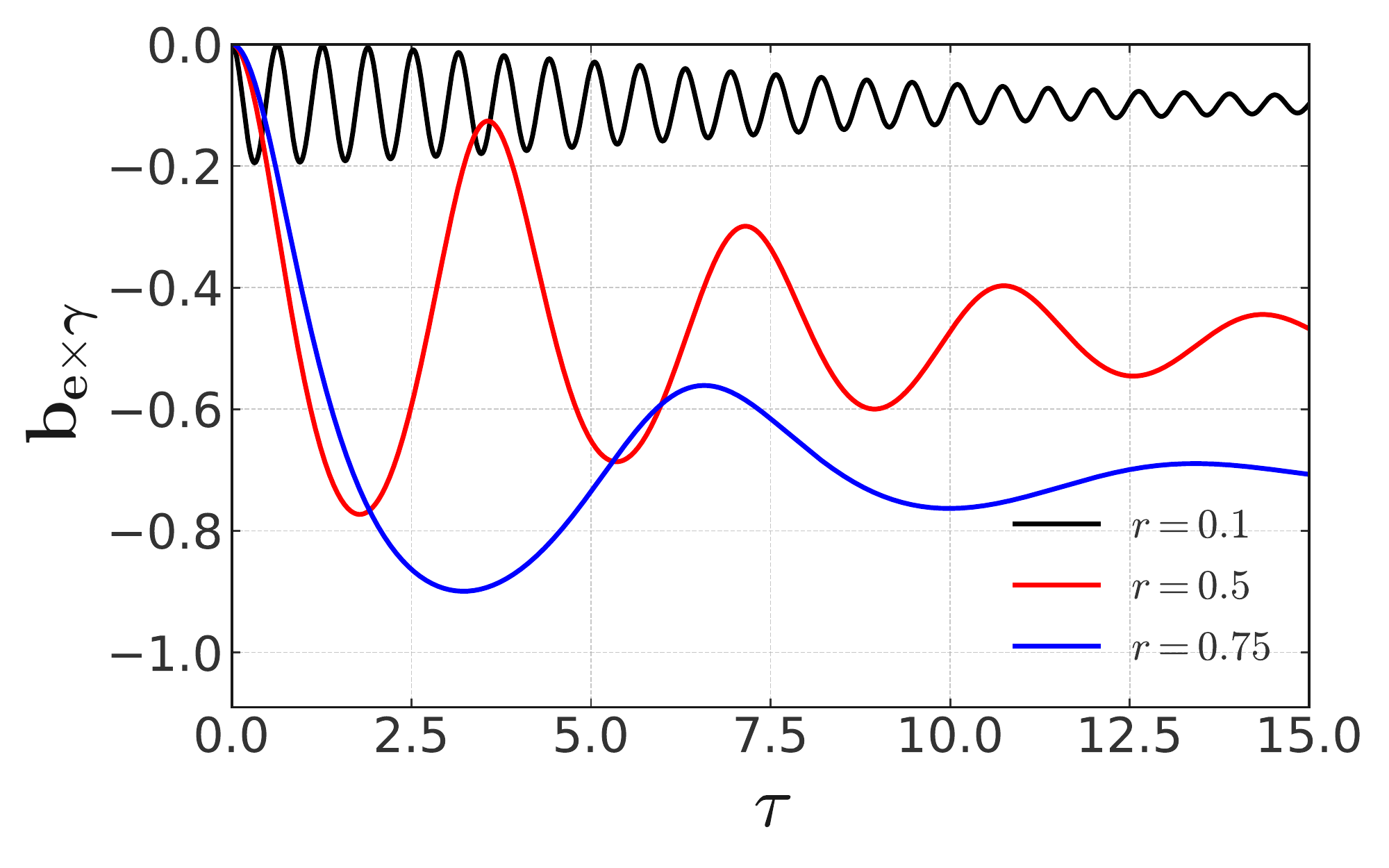}
    \caption{}
    \label{fig:rEnvelopeB}
    \end{subfigure}
    \caption{(a) The magnitude of the co-decaying Bloch vector, $\absb$. (b) Projection of the co-decaying Bloch vector in the direction $\mathbf{e} \times \boldsymbol{\gamma}$. We consider quasi-critical unstable qubits for three values of $r$, with $\theta_{\mathbf{e}\boldsymbol{\gamma}} = 80^\circ$, and the initial condition $\mathbf{b}(0) = \mathbf{0}$.}
    \label{fig:rEnvelope}
\end{figure}
\begin{figure}[t!]
    \centering
    \begin{subfigure}[t]{0.49\textwidth}
    \centering    
    \includegraphics[width=\linewidth]{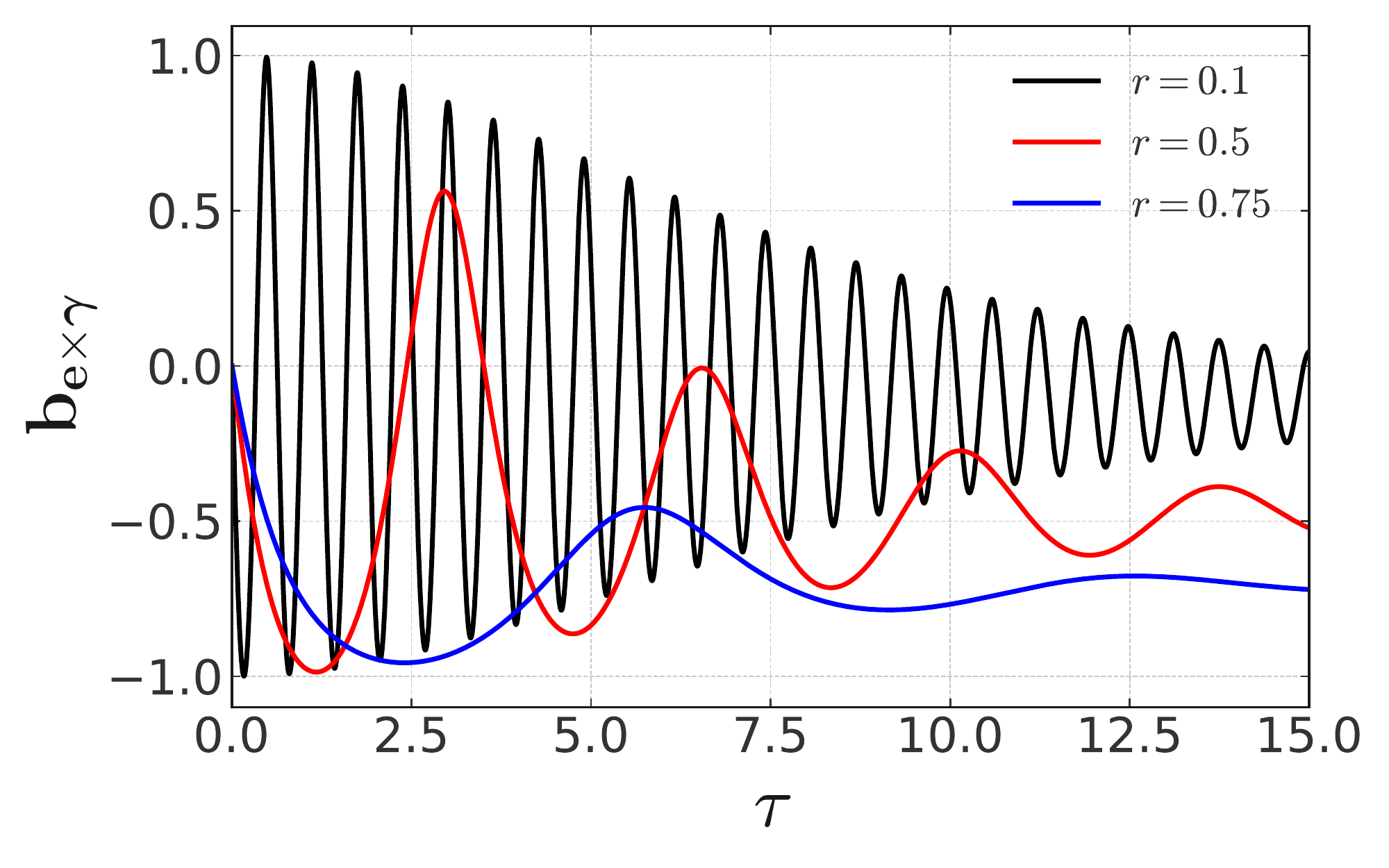}
    \caption{}
    \label{fig:CohrEnvelope}
    \end{subfigure}
    \begin{subfigure}[t]{0.49\textwidth}
    \centering
    \includegraphics[width=\linewidth]{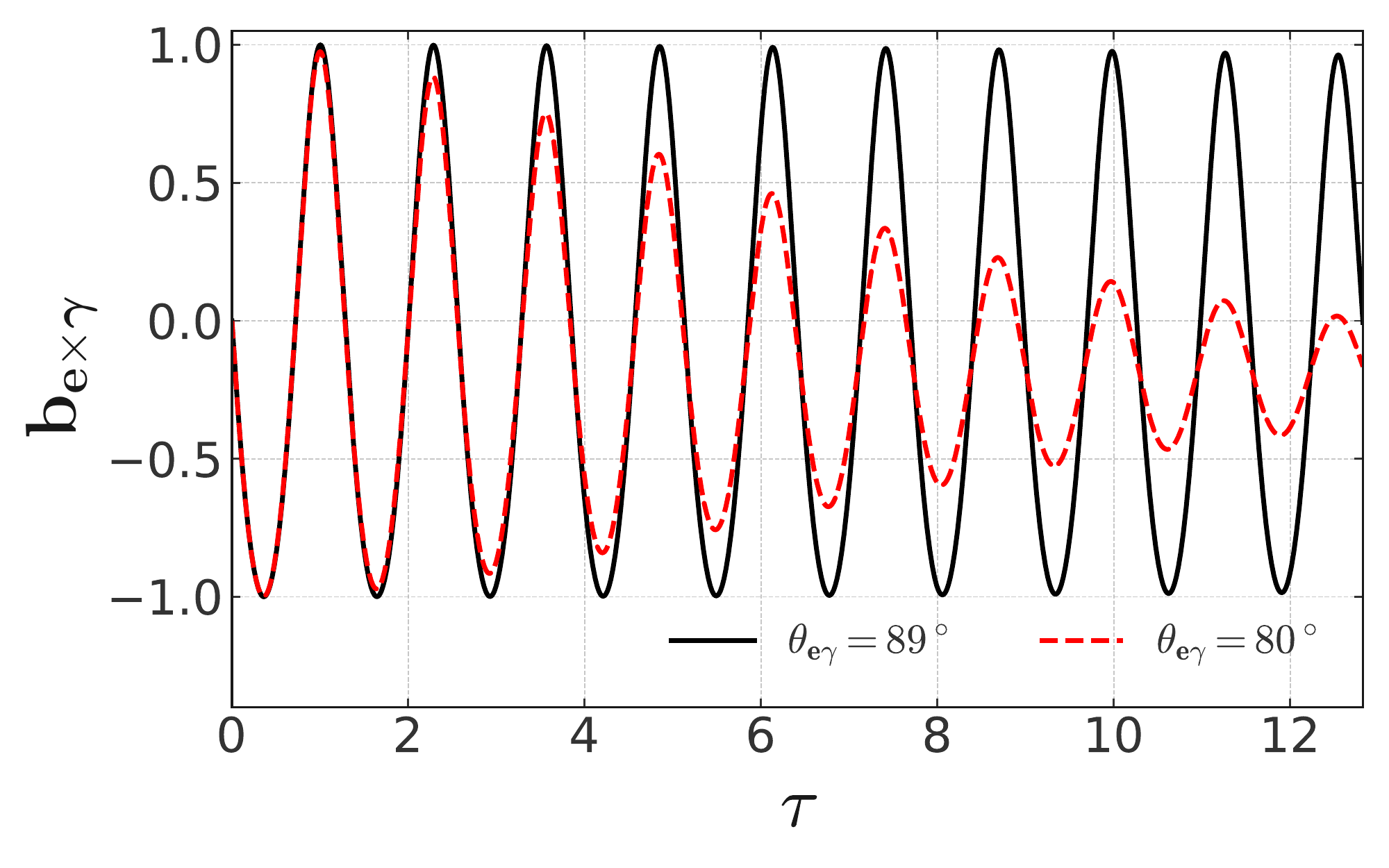}
    \caption{}
    \label{fig:CohThetaEnvelope}
    \end{subfigure}
    \caption{(a) Projection of the co-decaying Bloch vector in the direction $\mathbf{e} \times \boldsymbol{\gamma}$ for three values of $r$, but $\theta_{\mathbf{e}\boldsymbol{\gamma}}$ constant at $80^\circ$. (b) Projection of the co-decaying Bloch vector in the direction $\mathbf{e} \times \boldsymbol{\gamma}$ for two values of $\theta_{\mathbf{e}\boldsymbol{\gamma}}$, with $r=0.2$. We assume the initial condition $\mathbf{b}(0) = \boldsymbol{\gamma}$.}
    \label{fig:CohEnvelope}
\end{figure}
Figure~\ref{fig:ThetaEnvelope} shows the change in the behaviour of the co-decaying Bloch vector when the system is away from the critical scenario. To highlight this change, we exhibit both a small deviation away from the critical scenario, with $\theta_{\mathbf{e}\boldsymbol{\gamma}} = 89^\circ$, as well as a larger deviation, where $\theta_{\mathbf{e}\boldsymbol{\gamma}} = 80^\circ$. Figure~\ref{fig:rEnvelope} builds upon this, by varying the value of $r$ with $\theta_{\mathbf{e}\boldsymbol{\gamma}}$ fixed to $80^\circ$. Three different scenarios for quasi-critical unstable qubits are chosen to represent large, small and intermediate values of $r$, while assuming a fully mixed initial state ($\mathbf{b}(0) = \mathbf{0}$). In Figure~\ref{fig:CohEnvelope}, we analyse the same same set of scenarios, but with the assumption of a pure initial state, with~$|\mathbf{b}(0)| = 1$.

Figure~\ref{fig:ThetaEnvelope} demonstrates that any deviation away from the critical scenario results in a general tendency toward a fixed pure state. However, in Figure~\ref{fig:ThetaEnvelopeA}, one can see that the features of the critical scenario still pervade. In particular, we see that in quasi-critical scenarios, unstable qubits may exhibit 
damped coherence-decoherence oscillations, where the damping factor becomes smaller in systems closer to the critical scenario. Furthermore, as can be seen in Figure~\ref{fig:ThetaEnvelopeB}, small variations in the angle, $\theta_{\mathbf{e}\boldsymbol{\gamma}}$, do not alter the oscillation period, but only the rate at which the qubit purifies.

Nevertheless, Figure~\ref{fig:rEnvelope} shows that the angular separation between $\mathbf{e}$ and $\boldsymbol{\gamma}$ is not sufficient to provide by itself an accurate description of the time evolution of ${\bf b}(\tau )$. By considering different values of $r$, it is clear that very different behaviour may be observed. In particular, the period of coherence-decoherence oscillations is increased for larger values of $r$, and the qubit approaches a pure state earlier. Therefore, by calculating the oscillation period in the co-decaying frame, it is possible to find the value for~$r$.

In Figure~\ref{fig:CohEnvelope}, projections along the $\mathbf{e}\times\boldsymbol{\gamma}$ axis are presented in scenarios with $|\mathbf{b}(0)| = 1$. As was discussed in Section~\ref{sec:2Level}, such scenarios preserve the purity of the system ($|\mathbf{b}| = 1$) throughout their time evolution. As a consequence, the initial oscillations along this axis occur with unit amplitude. However, in quasi-critical scenarios, due to their non-zero torsion $\mathcal{T}$ as defined in~\eqref{eq:Torsion}, the Bloch vector moves off the $\left(\boldsymbol{\gamma}, \mathbf{e}\times\boldsymbol{\gamma} \right)$-plane, whilst remaining pure with $|\mathbf{b}| = 1$. Accordingly, the oscillation amplitude in the $\mathbf{e}\times\boldsymbol{\gamma}$ axis is reduced, and the qubit approaches its longest lived state. The manner in which this occurs is dependent on the values of $r$ and $\theta_{\mathbf{e}\boldsymbol{\gamma}}$. In Figure~\ref{fig:CohrEnvelope}, we vary the value of $r$ whilst retaining $\theta_{\mathbf{e}\boldsymbol{\gamma}} = 80^\circ$. As a result, we see that increasing the value of $r$ produces oscillations with smaller amplitude, but longer period, see (\ref{eq:A=0_crit}). Alternatively, we may vary $\theta_{\mathbf{e}\boldsymbol{\gamma}}$, whilst keeping $r=0.2$ constant, as displayed in Figure~\ref{fig:CohThetaEnvelope}. Here, it may be seen that maintaining the value of $r$ produces oscillations of identical period. However, larger departures from the critical scenario push the qubit to its long-lived state at an enhanced rate.

From \Figs{fig:ThetaEnvelope,fig:rEnvelope,fig:CohEnvelope}, we observe that the period of oscillation, $\widehat{\text{P}}$, gives us a significant insight into the value of $r$. In particular, we may deduce the value of $\theta_{\mathbf{e}\boldsymbol{\gamma}}$ through the value of $\mathbf{b}_\star$ or by considering the rate at which the unstable qubit reaches purity. Hence, precision measurement of the oscillation period and the asymptotic value of $\mathbf{b}$ in quasi-critical scenarios may be used to probe the structure of the effective Hamiltonian.

\subsection{CP Violating Quasi-Critical Scenarios}

In quasi-critical cases, with $|\cos\theta| \ll 1$ and $r<1$, we can approximate the CP-violating parameters \eqs{eq:pq_ratio,eq:mass_width_diffs} as follows:
\begin{equation}    
    \Delta\Gamma\: \simeq\: -\,4\absE \, \dfrac{r}{\sqrt{1-r^2}} \, \cos \theta_{\mathbf{e}\boldsymbol{\gamma}} \, , \qquad
    \left| \frac{q}{p} \right|\: \simeq\: \sqrt{1 - 2\dfrac{r}{1+r^2}\,\sin \theta_{\mathbf{e}\boldsymbol{\gamma}}} \;,
    \label{eq:meson_params_semi_crit}
\end{equation}
with $\Delta m$ given in \eqs{eq:meson_params_perp}. As shown in Table~\ref{tab:MesonData}, such scenario can be realised in the $B_{\rm d}^0$ meson system, as $\Gamma$ is allowed to be not exactly perpendicular to $\vecE$.  

Considering an initial fully mixed state, with $\vecb(0) = \pmb{0}$, enables us to express the magnitude of the Bloch vector as follows:
\begin{equation}
    \absb^2\ \simeq\ 1\: -\:  \lrb{1-r^2}^2\, \bigg[\cosh\lrb{ \dfrac{\cot \theta_{\mathbf{e}\boldsymbol{\gamma}}}{\sqrt{1-r^2}} \, \tau }  -r^2 \, \cos \lrb{\dfrac{\sqrt{1-r^2}}{r} \,\tau }\bigg]^{-2}
    \;,
    \label{eq:b_semi_crit}
\end{equation}
which reduces to \eqs{eq:b_crit} at $\theta = \pm \pi/2$. Observe that the co-decaying Bloch vector describes an oscillation with a decaying amplitude and a maximum value that asymptotically approaches $|\vecb_{\star}| = 1$. The $\tau$-evolution of $\absb$ in this case can be also seen in \Figs{fig:ThetaEnvelopeA,fig:rEnvelopeA}.
Note that, as explained in Section~\ref{sec:Meson}, a non-vanishing $\Delta \Gamma$ implies that the system asymptotically tends to the eigenstate of effective Hamiltonian ${\rm H}_{\rm eff}$ which has the longest lifetime with $|\vecb_{\star}|=1$. 


\section{Conclusions}\label{sec:Concl}
\setcounter{equation}{0}

We performed a detailed analysis of the dynamics of unstable two-level quantum systems within the context of the Bloch-sphere formalism. In our analysis, we used the Bloch-vector representation in order to describe such unstable qubit systems, by decomposing the effective Hamiltonian of the system in terms of the energy-level and decay-width vectors, ${\bf E}$ and~${\bf\Gamma}$, respectively. In general, at sufficiently large times, an initially mixed unstable qubit, which has a vanishing co-decaying Bloch vector ${\bf b}(\tau)$ at $\tau =0$, will tend to a pure state along its long-lived eigenstate, since its complementary short-lived eigenstate has already decayed away. However, for a system in which the vectors ${\bf E}$ and ${\bf\Gamma}$ are orthogonal to one another and the ratio $r = |{\bf \Gamma}|/(2|{\bf E}|)$ is less than~1, both eigenstates of the effective Hamiltonian have exactly equal lifetimes, and their large-time dynamics gets more complex.  They define, as we have called in this paper, a {\em critical} unstable two-level quantum system, or a {\em critical unstable qubit}. 

We have found for the first time that critical unstable qubit systems being initially at a fully mixed state (with ${\bf b}(0) = {\bf 0}$) display an unusual coherence--decoherence oscillation behaviour when studied in a conveniently defined co-decaying frame of the system. We must stress here that these oscillations are novel phenomena and emerge beyond the usual oscillatory pattern due to the energy-level difference of the qubit. Interestingly enough, we have demonstrated how these new features persist even for quasi-critical scenarios, in which the vectors~${\bf E}$ and ${\bf\Gamma}$ happen to be not perfectly orthogonal to one another. As mentioned above, if a generic unstable qubit is initially prepared as a pure state characterised by a co-decaying Bloch vector ${\bf b}$ with norm~1 at~${\tau = 0}$, it will then  remain so at all times and the unit vector ${\bf b}(\tau )$ will eventually freeze along the direction of the long-lived eigenstate of the two-level system. However, for critical unstable qubits, the situation turns out to be very different. In this case, we observe that the unit Bloch vector ${\bf b}(\tau)$ of the qubit will rotate about the direction defined by the energy-level vector~${\bf E}$, and it will sweep out {\em unequal} areas in equal times. This is in stark contradistinction with the behaviour of ${\bf b}(\tau)$ for a stable or uncritical qubit, and amusingly enough, with Kepler's second law of plenatary motion. In this context, we have shown that for any non-trivial critical unstable two-level system, the Bloch vector ${\bf b}(\tau )$ describes a trajectory that lies always on a plane orthogonal to the energy-level vector~${\bf E}$. 

In addition, we have studied the impact of decoherence phenomena, as described by the decoherence vector ${\bf D}$, on the time evolution of the Bloch vector ${\bf b}(\tau )$ in both critical and quasi-critical scenarios. In all cases of interest, the Bloch vector ${\bf b}(\tau)$ will freeze in a direction that can be fully specified in terms of the decoherence parameter $\zeta = 4|{\bf D}|^2/|{\bf \Gamma}|$ and the unit vector~${\bf d} = {\bf D}/|{\bf D}|$.  Evidently, critical unstable qubits 
exhibit unusual quantum behaviours that have no analogue in classical systems. The simulation of their critical phenomena might require the use of quantum computers to probe the validity of the many unexpected predictions presented in this work.
Likewise, one may be tempted to go beyond the $d\!=\!2$ level case that we have been studying here, and analyse critical phenomena in unstable $d$--level quantum systems,  especially when $d$ is a prime number, like $d=3,5$, in which the dynamics of the so-called unstable {\em qudit} system becomes irreducible.

In conclusion, we note that utilising the co-decaying Bloch vector ${\bf b}(\tau)$ within experimental setups can offer new vistas in probing the structure of the effective Hamiltonian for unstable quantum systems. In this context, we have discussed potential applications of our results to CP-violating unstable meson--antimeson systems formed by $K$-, $B$- and $D$-mesons. In particular, we have found that the $B^0_d\bar{B}^0_d$ system might realise a critical two-level system, and future experiments are deemed crucial to shed light on its properties. 
However, we believe that the Bloch-sphere approach developed here can play a pivotal role in the analysis of many different unstable qubit systems in various fields of particle and astro-particle physics, like axion-photon mixing, sterile-neutrino oscillations, or in describing the muon precession in the $g-2$ experiment at Fermilab.

\subsection*{Acknowledgements} 
\vspace{-1mm}

\noindent
The work of AP and DK is supported in part by the Lancaster-Manchester-Sheffield Consortium for Fundamental Physics, under STFC Research Grant ST/T001038/1. TM acknowledges support from the STFC Doctoral Training Partnership under STFC training grant ST/V506898/1.

\newpage
\setcounter{section}{0}
\section*{Appendix}
\appendix

\renewcommand{\theequation}{\Alph{section}.\arabic{equation}}
\setcounter{equation}{0}  
\section{Analytic Results}\label{app:Analytic}

The system \eqs{eq:a0_eom,eq:ai_eom} can be solved by evolving the density matrix from its initial state using \eqs{eq:rho_evolution_operator}. Then, the Bloch 4-vector $a^\mu$ can be obtained using the identity $a^\mu  = \Tr{\sigma^\mu \, \rho}$, which yields
\begin{align}
        a^0 =& \dfrac{e^{- \Gamma^0 \, t}}{\absH^*\absH}\lrBigcb{
        \absH^*\absH \, \cos\lrb{ \absH^* \, t  } \ \cos\lrb{ \absH \, t  }  \nonumber\\
        &-8\,\Re{ \absH \lrb{\veca_0 \cdot \vecH^*} \ \cos\lrb{ \absH^* \, t  } \ \sin\lrb{ \absH \, t  }   }\nonumber\\
        & -4\lrb{\absE^2 \, (1+r^2) + \veca_0 \cdot \lrb{\vecE \times \vecG}} \ \sin\lrb{ \absH^* \, t  } \ \sin\lrb{ \absH \, t  }
        }     
    %
    \label{eq:a_solution} \\
    %
        \veca =& \dfrac{e^{- \Gamma^0 \, t}}{\absH^*\absH}\lrBigcb{
        \veca_0 \, \absH^*\absH \, \cos\lrb{ \absH^* \, t  } \ \cos\lrb{ \absH \, t  }  \nonumber\\
        &+2\, \Re{ \absH^*\lrb{i \vecH + \veca_0 \times \vecH }\, \cos\lrb{ \absH^* \, t  } \ \sin\lrb{ \absH \, t  }}\nonumber\\
        &-2 \lrsb{ \lrb{\veca_0 \cdot \vecG} \, \vecG + 4\lrb{\veca_0 \cdot \vecE} \, \vecE  -2 \lrb{\vecH \cdot \vecH^*}  \, \veca_0 + 2 \vecE \times \vecG}
        \sin\lrb{ \absH^* \, t  } \ \sin\lrb{ \absH \, t  }
        }\;,
    %
    \nonumber
\end{align}
where $\absH = \sqrt{\vecH \cdot \vecH} \, \in\,\mathbb{C} $ and the initial condition for $\veca$ is denoted as $\veca_0$. The co-decaying Bloch vector can be obtained by its definition, $\vecb = \veca/a^0$. 

\section{Non-Linear Analysis}\label{app:Nonlinear}
\setcounter{equation}{0}

The location of fixed points in non-linear systems is well known to be characterised by the vanishing of the equation of motion. However, an assessment of the stability for these fixed points is a far less trivial exercise. In this appendix, we outline a method, by which one may identify the stable fixed points for the non-linear system of equations that we consider in~\eqref{eq:dbdt}.

To this end, we construct the right-handed orthonormal basis spanned by the unit vectors~$\mathbf{e}$ and~$\boldsymbol{\gamma}$  as follows:
\begin{equation}
    \mathbf{n}_1 = \mathbf{e}, \qquad \mathbf{n}_2 = \frac{1}{s_\gamma} \mathbf{e} \times \boldsymbol{\gamma}, \qquad \mathbf{n}_3 = \frac{1}{s_\gamma} \mathbf{e} \times (\mathbf{e} \times \boldsymbol{\gamma}),
\end{equation}
where $s_\gamma = \sin \theta_{\mathbf{e}\boldsymbol{\gamma}}$ is the sine of the angle between the unit vectors $\mathbf{e}$ and $\boldsymbol{\gamma}$.
After rewriting the equation of motion in this new basis, we may consider an ansatz for the solution of the form,
\begin{equation}
    \mathbf{b} = \alpha \mathbf{n}_1 + \beta \mathbf{n}_2 + \eta \mathbf{n}_3,
\end{equation}
and derive three coupled autonomous equations for the evolution of the components of $\mathbf{b}$
\begin{subequations}
\begin{align}
    \frac{d\alpha}{d\tau} &= c_\gamma - \left( c_\gamma \alpha - s_\gamma \eta \right)\alpha,\\
    \frac{d\beta}{d\tau} &= \frac{1}{r}\eta - \left(c_\gamma \alpha - s_\gamma \eta \right)\beta,\\
    \frac{d\eta}{d\tau} &= -\frac{1}{r}\beta -s_\gamma - \left(c_\gamma \alpha - s_\gamma \eta \right)\eta,
\end{align}
\end{subequations}
with $c_\gamma = \cos \theta_{\mathbf{e}\boldsymbol{\gamma}}$. We may determined the fixed points of the system by simultaneously taking each derivative to zero \cite{Marsden:1976}, $\alpha^\prime = \beta^\prime = \eta^\prime = 0$. Solving the resulting system of equations gives
\begin{equation}
    \beta = - \frac{1}{s_\gamma r} (1-\alpha^2), \qquad \eta = - \frac{c_\gamma}{s_\gamma \alpha} (1-\alpha^2)\,,
\end{equation}
as well as the polynomial constraint
\begin{equation}
    \alpha^4 - (1-r^2)\alpha^2 - c_\gamma^2 r^2 = 0\,.
\end{equation}
This polynomial of course has four solutions, although we are only interested in the solutions which correspond to asymptotic fixed points. Of all possible solutions, two may be ruled out by requiring that the co-decaying Bloch vector $\vecb$ is real ($\alpha^2 \geq 0$), implying that
\begin{equation}
    \alpha\: =\: \pm \frac{1}{\sqrt{2}}\, \sqrt{1-r^2 + \sqrt{(1-r^2)^2 + 4c_\gamma^2 r^2}}\; .
\end{equation}
Whether we should take the positive or negative root is highly non-trivial. since we must assess whether the fixed points they describe are stable. The stability of a dynamical system is characterised by the eigenvalues of its Jacobian at the fixed point,
\begin{equation}\label{eq:Jacobian}
    \mathbb{J} = \left( \begin{matrix} 
    - \frac{c_\gamma}{\alpha}(1+\alpha^2) & 0 & s_\gamma \alpha \\
    \frac{c_\gamma}{s_\gamma r}\left(1-\alpha^2 \right) & -\frac{c_\gamma}{\alpha} & \frac{1}{r}\alpha^2 \\
    \frac{c_\gamma^2}{s_\gamma \alpha}(1-\alpha^2) & -\frac{1}{r} & -\frac{c_\gamma}{\alpha}(2-\alpha^2)
    \end{matrix} \right).
\end{equation}
For such points to be stable, it is necessary that the real part of all eigenvalues be negative \cite{Marsden:1976}. However, finding the eigenvalues of such a complicated system is in general rather difficult. Instead, we will use the general properties of the Jacobian to infer relevant information. Since we are searching for negative eigenvalues, and the Jacobian is of odd rank, we know that $\textrm{Det}\,\mathbb{J} < 0$, and $\textrm{Tr}\,\mathbb{J} < 0$ are necessary conditions. Calculating these quantities yields the following expressions:
\begin{equation}\label{eq:FPCond}
    \textrm{Det}\,\mathbb{J}\: =\: - \left( \frac{c_\gamma}{\alpha} \right)^3 \left[ 1 + \frac{\alpha^4}{c_\gamma^2 r^2}\right], \qquad \textrm{Tr}\,\mathbb{J}\: =\: -4 \frac{c_\gamma}{\alpha}.
\end{equation}
Therefore, in order to guarantee that the system corresponds to a fixed point, the branch we take for $\alpha$ must be of the same sign as $c_\gamma$. In this way, we obtain the asymptotic value~${\bf b}_\star$ of the co-decaying Bloch vector in terms of the input parameters, as given in (\ref{eq:FPN}).

\newpage 

\bibliography{bibs-refs}{}
\bibliographystyle{JHEP}
\end{document}